\pdfoutput=1

\documentclass[11pt,twoside,a4paper,cmspaper,final,collab]{cms-tdr}
\def\svnVersion{405625}\def\cmsCernNoTag{CERN-EP/2016-284}\def\cmsCernDate{\today}\def\cmsMessage{Published in the European Physical Journal C as \href{http://dx.doi.org/10.1140/epjc/s10052-017-4853-2}{\doi{10.1140/epjc/s10052-017-4853-2.}}}

\begin{document}\cmsNoteHeader{SUS-16-008}

\hyphenation{had-ron-i-za-tion}
\hyphenation{cal-or-i-me-ter}
\hyphenation{de-vices}
\RCS$Revision: 405598 $
\RCS$HeadURL: svn+ssh://svn.cern.ch/reps/tdr2/papers/SUS-16-008/trunk/SUS-16-008.tex $
\RCS$Id: SUS-16-008.tex 405598 2017-05-19 15:19:39Z peveraer $
\newlength\cmsFigWidth
\ifthenelse{\boolean{cms@external}}{\setlength\cmsFigWidth{0.98\columnwidth}}{\setlength\cmsFigWidth{0.8\textwidth}}
\ifthenelse{\boolean{cms@external}}{\providecommand{\cmsLeft}{top\xspace}}{\providecommand{\cmsLeft}{left\xspace}}
\ifthenelse{\boolean{cms@external}}{\providecommand{\cmsRight}{bottom\xspace}}{\providecommand{\cmsRight}{right\xspace}}
\ifthenelse{\boolean{cms@external}}{\providecommand{\cmsTable}{\relax}}{\providecommand{\cmsTable}[1]{\resizebox{0.7\textwidth}{!}{#1}}}
\providecommand{\NA}{\ensuremath{\text{---}}}
\newcommand{\relmu}{\mbox{Re} (\hat{\mu}_{\PQt})}
\newcommand{\imd}{\mbox{Im} (\hat{d}_{\PQt})}
\newcommand{\ttfake}{\ttbar\ (non-dileptonic)}
\newcommand{\MTtW}{\ensuremath{M_{\mathrm{T2}}^{\PW}}\xspace}
\newcommand{\HTonetwo}{\ensuremath{H_{\mathrm{T,12}}}}
\newcommand{\mtop}{\ensuremath{m_\PQt}\xspace}
\newcommand{\ttll}{\ensuremath{\ttbar\to\ell\ell}\xspace}
\newcommand{\ttlj}{\ensuremath{\ttbar\to\ell+\text{jets}}\xspace}
\newcommand{\pp}{\ensuremath{\Pp\Pp}\xspace}
\newcommand{\ppbar}{\ensuremath{\Pp\Pap}\xspace}
\newcommand{\eepm}{\ensuremath{\Pep\Pem}\xspace}
\newcommand{\mmpm}{\ensuremath{\Pgmp \Pgmm}\xspace}
\newcommand{\ttpm}{\ensuremath{\Pgt^+ \Pgt^-}\xspace}
\newcommand{\empm}{\ensuremath{\Pe^\pm \Pgm^\mp}\xspace}
\providecommand{\Pepm}{\ensuremath{\Pe^\pm}\xspace}

\newcommand{\nb}{\ensuremath{N_{\PQb}}\xspace}
\newcommand{\nj}{\ensuremath{N_{\mathrm{j}}}\xspace}
\newcommand{\nt}{\ensuremath{N_{\PQt}}\xspace}
\newcommand{\met}{\MET}
\newcommand{\metg}{\ensuremath{E^{\text{miss},\gamma}_{\mathrm{T}}}\xspace}
\newcommand{\metll}{\ensuremath{E^{\text{miss},\ell\ell}_{\mathrm{T}}}\xspace}
\newcommand{\tfllb}{\ensuremath{T_{\text{LL}}}\xspace}
\newcommand{\znunu}{\ensuremath{\cPZ\to\cPgn\cPagn}\xspace}
\newcommand{\znunuM}{\ensuremath{\cPZ\to\cPgn\cPagn}\xspace}
\newcommand{\zll}{\ensuremath{\cPZ\to\ell\ell}}
\newcommand{\W}{\ensuremath{\PW}\xspace}
\newcommand{\ffbar}{\ensuremath{\cmsSymbolFace{f}\overline{\cmsSymbolFace{f}}}\xspace}
\newcommand{\ttbarZ}{\ensuremath{\ttbar\cPZ}\xspace}
\newcommand{\ttbarW}{\ensuremath{\ttbar\PW}\xspace}
\newcommand{\mtb}{\ensuremath{M_{\mathrm{T}}(\PQb_{1,2},\ptvecmiss)}\xspace}
\newcommand{\mtl}{\ensuremath{M_{\mathrm{T}}(\ell,\ptvecmiss)}\xspace}
\newcommand{\dphijonetwothree}{\ensuremath{\Delta\phi_{123}}\xspace}
\newcommand{\dphijonetwothreefour}{\ensuremath{\Delta\phi_{1234}}\xspace}
\newcommand{\stopq}{\ensuremath{\PSQt_{1}}\xspace}
\newcommand{\stopqbar}{\ensuremath{\PASQt_{1}}\xspace}
\newcommand{\sbottomq}{\ensuremath{\PSQb_{1}}\xspace}
\newcommand{\sbottomqbar}{\ensuremath{\PASQb_{1}}\xspace}
\newcommand{\topq}{\PQt}
\newcommand{\topqbar}{\cPaqt}
\newcommand{\bq}{\PQb}
\newcommand{\bqbar}{\cPaqb}
\newcommand{\cq}{\cPqc}
\newcommand{\cqbar}{\cPaqc}
\newcommand{\lsp}{\PSGczDo}
\newcommand{\chgp}{\PSGcpDo}
\newcommand{\chgm}{\PSGcmDo}
\newcommand{\chipmone}{\PSGcpmDo}
\newcommand{\rjet}{\ensuremath{r_{\text{jet}}}\xspace}
\newcommand{\prjet}{\ensuremath{r^{\mathrm{pseudo}}_{\text{jet}}}\xspace}
\newcommand{\TQCD}{\ensuremath{T_{\mathrm{QCD}}}\xspace}
\newcommand{\MT}{\ensuremath{M_{\mathrm{T}}}\xspace}
\newcommand{\mct}{\ensuremath{m_{\mathrm{CT}}}\xspace}
\newcommand{\dphijmet}{\ensuremath{\Delta\phi(\mathrm{j}_{1},\ptvecmiss)}\xspace}
\newcommand{\minMT}{\ensuremath{M_{\mathrm{T}}(\mathrm{j}_{1,2},\ptvecmiss)}\xspace}
\newcommand{\zmm}{\ensuremath{\cPZ\to\mu^{+}\mu^{-}}\xspace}
\newcommand{\wjets}{\ensuremath{\PW+}jets\xspace}
\newcommand{\zjets}{\ensuremath{\Z+}jets\xspace}
\newcommand{\gjets}{\ensuremath{\gamma+}jets\xspace}

\cmsNoteHeader{SUS-16-008}
\title{Searches for pair production of third-generation squarks in $\sqrt{s}=13$\TeV pp collisions}

\date{\today}

\abstract{ Searches are presented for direct production of top or bottom squark pairs in proton-proton collisions at the CERN LHC. Two searches, based on complementary techniques, are performed in all-jet final states that are characterized by a significant imbalance in transverse momentum. An additional search requires the presence of a charged lepton isolated from other activity in the event. The data were collected in 2015 at a centre-of-mass energy of 13\TeV with the CMS detector and correspond to an integrated luminosity of 2.3\fbinv. No statistically significant excess of events is found beyond the expected contribution from standard model processes. Exclusion limits are set in the context of simplified models of top or bottom squark pair production. Models with top and bottom squark masses up to 830 and 890\GeV, respectively, are probed for light neutralinos. For models with top squark masses of 675\GeV, neutralino masses up to 260\GeV are excluded at 95\% confidence level.
}

\hypersetup{%
pdfauthor={CMS Collaboration},%
pdftitle={Searches for pair production of third-generation squarks in sqrt(s)=13 TeV pp collisions},%
pdfsubject={CMS},%
pdfkeywords={CMS, physics, SUSY}}

\maketitle

\section{Introduction}
\label{sec:intro}
The standard model (SM) has been extremely successful at describing particle physics phenomena. Nevertheless, it suffers from shortcomings such as the hierarchy problem~\cite{'tHooft:1979bh,Witten:1981nf,Dine:1981za,Dimopoulos:1981au,Dimopoulos:1981zb,Kaul:1981hi}, the need for fine-tuned cancellations of large quantum corrections to keep the Higgs boson mass near the electroweak scale.
Supersymmetry (SUSY), based on a symmetry between bosons and fermions, is an attractive extension of the SM. A key feature of SUSY is the existence of a superpartner for every SM particle with the same quantum numbers, except for spin, which differs by one half unit. In R-parity conserving SUSY models~\cite{Wess:1974tw,Farrar:1978xj}, supersymmetric particles are created in pairs, and the lightest supersymmetric particle (LSP) is stable~\cite{djoua,carena} and considered to be a candidate for dark matter~\cite{darkmatter}.
 Supersymmetry can potentially provide a ``natural'', \ie not fine-tuned, solution to the hierarchy problem through the cancellation of quadratic divergences in particle and sparticle loop corrections to the Higgs boson mass.
In natural SUSY models light top and bottom squarks with masses close to the electroweak scale are preferred.

This paper presents three complementary searches for direct production of a pair of top ($\stopq\stopqbar$) or bottom squarks ($\sbottomq\sbottomqbar$), where the subscript here denotes the less massive partner of the corresponding SM fermion's chirality states. The first search targets top squark pair production in the all-jet final state, while the second focuses on the single-lepton final state.
These two analyses were explicitly designed for complementarity, allowing for a combination of the results to enhance the sensitivity. The third search targets bottom squark pair production in the all-jet final state.
The searches are performed using the data collected in proton-proton collisions at a centre-of-mass energy of 13\TeV with the CMS detector at the CERN LHC in 2015, corresponding to an integrated luminosity of 2.3\fbinv .
The results of similar searches were previously reported by the ATLAS and CMS collaborations using proton-proton collisions at 7 and 8\TeV~\cite{ATLAS1,ATLAS2,ATLAS5,ATLAS5a,ATLAS6,ATLAS7,ATLAS8,atlas-stop0l-2014,atlas-stop1l-2015,CMS-STOP-lepton,CMS-alphaT,RAZOR_8TeV,stop8TeV,stop0l_8TeV} and by the CDF and D0 collaborations in $\Pp\bar{\Pp}$ collisions at 1.96\TeV at the Fermilab Tevatron~\cite{Abazov:2003wt,Abazov:2012cz,Abazov:2007ak,Aaltonen:2010uf,Acosta:2003ys}.
With the increase in LHC collision energy from 8 to 13\TeV, the cross section to produce signal events is enhanced by a factor of 8--12 for a top or bottom squark mass in the range 700--1000\GeV~\cite{Borschensky:2014cia,SUSYXCLHC}. Therefore, new territory can be explored even with the relatively small amount of data collected in 2015. The CMS and ATLAS collaborations have already provided first exclusion results for these models in the all-jet and single-lepton final states~\cite{CMS-PAS-SUS-15-003,CMS-PAS-SUS-15-004,CMS-PAS-SUS-15-005,atlas-stop1l-2016}. Unlike the more generic searches for new phenomena presented by the CMS collaboration in Refs.~\cite{CMS-PAS-SUS-15-003,CMS-PAS-SUS-15-004,CMS-PAS-SUS-15-005}, the searches described in this paper directly target top and bottom squark production through the design of search regions that exploit the specific characteristics of these signal models, for instance through the use of a top quark tagging algorithm in the top squark search in the all-jet final state to identify boosted hadronically decaying top quarks originating from top squark decays.

\begin{figure*}[!htpb]
\centering
\includegraphics[width=\textwidth]{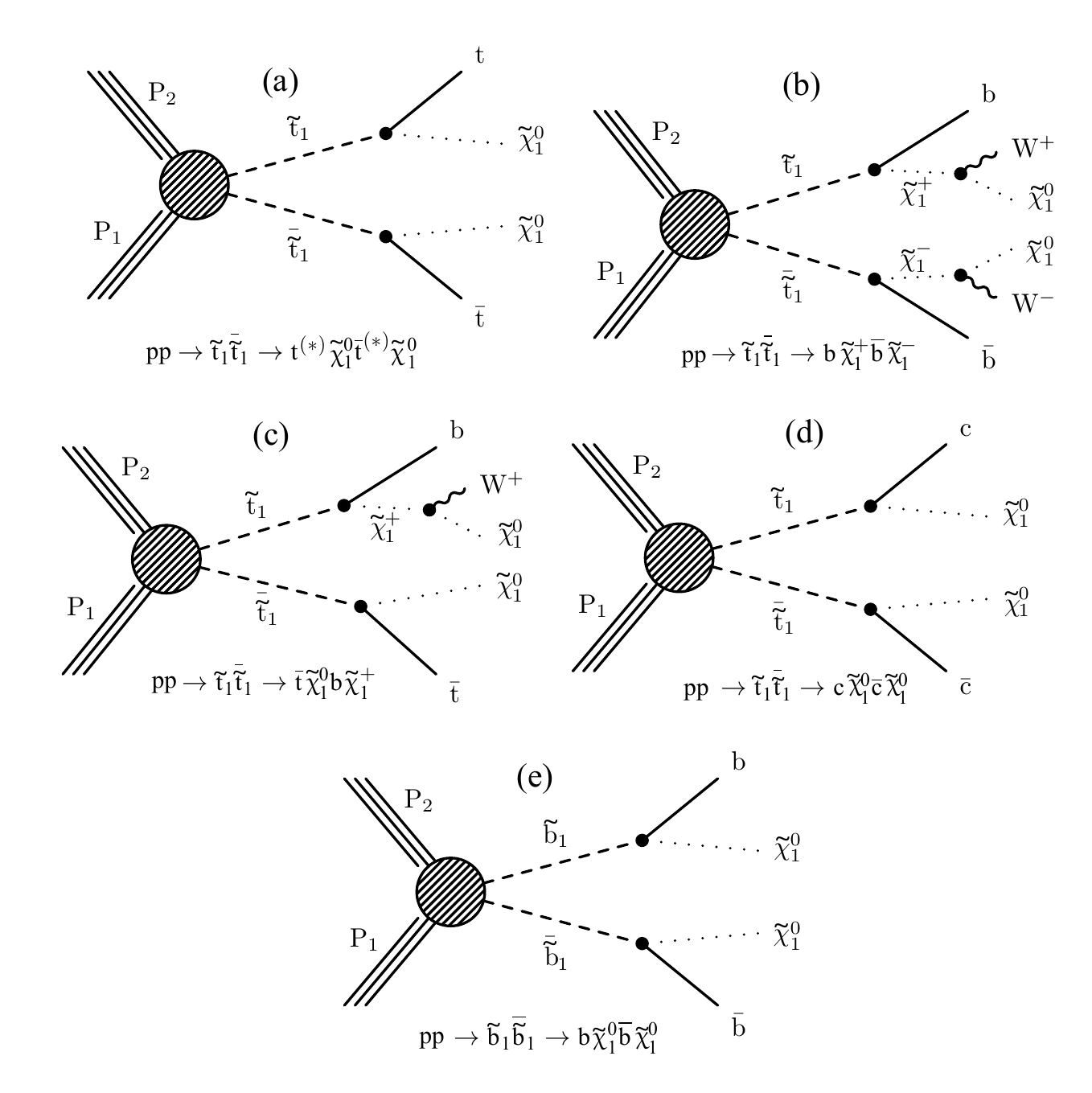}\caption{
  \label{fig:diagram}
        Feynman diagrams for pair production of top and bottom squarks via the decay modes considered in this paper.  The model with 50\% branching fractions for $\stopq \to \topq^{(*)} \lsp$ and $\stopq \to \bq \chipmone \to \bq \W^{\pm*} \lsp$ decays leads to the final states in diagrams (a)--(c).}
\end{figure*}

The decay modes of top squarks depend on the sparticle mass spectrum. Figure~\ref{fig:diagram} illustrates the top and bottom squark decay modes explored in this paper. The simplest top squark decay modes are $\stopq \to \topq^{(*)} \lsp$ and $\stopq \to \bq \chipmone \to \bq \W^{\pm(*)} \lsp$, with $\chipmone$ representing the lightest chargino, and with intermediate particles that can be virtual marked by asterisks. In these decay modes, the neutralino and charginos are mixtures of the superpartners of
electroweak gauge and Higgs bosons, and $\lsp$ is considered to be an LSP that escapes detection, leading to a potentially large transverse momentum imbalance in the detector.
The two analyses of top squark pair production in the all-jet and single-lepton final states probe both of these $\stopq$ decay modes.
In the $\stopq \to \topq^{(*)} \lsp$ decay mode, the top quark is produced off-shell when $\Delta m \equiv m_{\stopq}-m_{\lsp} < m_{\topq}$, while in the  $\stopq \to \bq \chipmone$ decay mode, the experimental signature is affected by the mass of the chargino.
We consider a model in which both top squarks decay via the $\stopq \to \topq^{(*)} \lsp$ decay mode. A second model in which the branching fraction for each of the two top squark decay modes is 50\% is also considered, under the assumption of a compressed mass spectrum in which the mass of $\chipmone$ is only 5\GeV greater than that of $\lsp$, with the \W~bosons resulting from chargino decays consequently being produced heavily off-shell.
If $\Delta m <m_{\W}$, $\stopq$ can decay through a four-body decay involving an SM fermion pair \ffbar as $\stopq\to\bq\ffbar\lsp$, or through a flavour changing neutral current decay $\stopq\to\cq\lsp$. The analysis of bottom squark pair production considers the decay mode $\sbottomq\to\bq\lsp$ within the allowed phase space, and also probes top squark pair production in the $\stopq\to\cq\lsp$ decay scenario.

This paper is organized as follows. Section~\ref{sec:detector} contains a brief description of the CMS detector, while Section~\ref{sec:obj} discusses the event reconstruction and simulation. Sections~\ref{sec:stop0l},~\ref{sec:1lstop}, and~\ref{sec:sbottom} present details for the all-jet top squark search, the single-lepton top squark search, and the all-jet bottom squark search, respectively.
Section~\ref{sec:systematics} describes the systematic uncertainties affecting the results of the three analyses. The interpretation of the results in the form of exclusion limits on models of top or bottom squark pair production is discussed in Section~\ref{sec:interpretation}, followed by a summary in Section~\ref{sec:summary}.
\section{The CMS detector}
\label{sec:detector}

The central feature of the CMS apparatus is a superconducting solenoid of 6\unit{m} internal diameter,
providing a magnetic field of 3.8\unit{T}. Within the solenoid volume are an all-silicon pixel and strip tracker, a
lead tungstate crystal electromagnetic calorimeter (ECAL), and a brass and scintillator hadron calorimeter (HCAL), each composed of
a barrel and two endcap sections.
Forward calorimeters extend the pseudorapidity ($\eta$) coverage provided by the barrel and endcap detectors.
 Muons are measured in gas-ionization detectors embedded in the steel flux-return yoke outside
the solenoid. The first level of the CMS trigger system, composed of custom hardware processors, uses information from the
calorimeters and muon detectors to select the most interesting events in a fixed time interval of less than 4\mus. The high-level
 trigger processor farm further decreases the event rate from around 100\unit{kHz} to around 1\unit{kHz}, before data storage.
A more detailed description of the CMS detector, together with a definition of the coordinate system used and the relevant
kinematic variables, can be found in Ref.~\cite{JINST}.

\section{Reconstruction algorithms and simulation}
\label{sec:obj}

Event reconstruction uses the particle-flow (PF) algorithm \cite{CMS:2009nxa,CMS:2010eua}, combining information from the tracker, calorimeter, and muon systems to identify charged hadrons, neutral hadrons, photons, electrons, and muons in an event. The missing transverse momentum, \ptvecmiss, is computed as the negative vector sum of the transverse momenta (\ptvec) of all PF candidates reconstructed in an event, and its magnitude \met is an important discriminator between signal and SM background. Events selected for the searches are required to pass filters designed to remove detector- and beam-related noise and must have at least one reconstructed vertex. Usually more than
one such vertex is reconstructed, due to pileup, i.e. multiple pp collisions within the same or neighbouring bunch crossings. The reconstructed vertex with the largest $\sum{\pt^2}$ of associated tracks is designated as the primary vertex.

Charged particles originating from the primary vertex, photons, and neutral hadrons are clustered into jets using the anti-\kt algorithm~\cite{Cacciari:2008gp} implemented in FastJet~\cite{Cacciari:2011ma} with a distance parameter of 0.4. The jet energy is corrected to account for the contribution of additional pileup interactions in an event and to compensate for variations in detector response~\cite{Cacciari:2011ma,pileup}. Jets considered in the searches are required to have their axes within the tracker volume, within the range $\abs{\eta} < 2.4$.

Jets originating from \bq~quarks are identified with the combined secondary vertex (CSV) algorithm~\cite{Chatrchyan:2012jua,CMS-PAS-BTV-15-001} using two different working points, referred to as ``loose" and ``medium". The \bq~tagging efficiency for jets originating from \bq~quarks is about 80\% and 60\% for the loose and medium working point, respectively, while the misidentification rates for jets from charm quarks, and from light quarks or gluons are about 45\%  and 12\%, and 10\% and 2\%, respectively.

The ``CMS top (quark) tagging'' (CTT) algorithm~\cite{CMS:2014fya,CMS:2016tvk,Kaplan:2008ie} is used to identify highly energetic top quarks decaying to jets with the help of observables related to jet substructure~\cite{Dasgupta:2013ihk,Cacciari:2005hq} and mass.
For a relativistic top quark with a Lorentz boost $\gamma = E/m$, the \W~boson and \bq~quark produced in the top quark decay are expected to be separated by a distance $R \equiv \sqrt{\smash[b]{(\Delta\eta)^2 + (\Delta\phi)^2} }\approx 2/\gamma$ (where $\phi$ is the azimuthal angle in radians).
In cases where the \W~boson subsequently decays hadronically, the three resulting jets from the \W~boson decay and the hadronization of the \bq~quark are likely to be merged into a single jet by a clustering algorithm with a distance parameter larger than $2/\gamma$.
To identify hadronically decaying top quarks with $\pt > 400$\GeV, we therefore use jets reconstructed using the anti-\kt algorithm with a distance parameter of 0.8 to try to cluster the top quark decay products into a single jet.
The next step of top quark reconstruction is an attempt to decompose the candidate jet into at least three subjets with the help of the Cambridge-Aachen jet clustering algorithm~\cite{Dokshitzer:1997in,CACluster2}, the invariant mass of which is required to be consistent with the top quark mass (140--250\GeV).  The final requirement of top quark identification is that the minimum invariant mass of any pair of the three subjets with the highest \pt must exceed 50\GeV.
The efficiency of the CTT algorithm to identify jets originating from top quark decays is measured to be about 30--40\% while the misidentification rate is found to be about 4--6\%, depending on the \pt of the top quark candidates. No disambiguation is performed between top quark candidates and jets reconstructed with a distance parameter of 0.4.

Electron candidates are reconstructed by first matching clusters of energy deposited in the ECAL to reconstructed tracks. Selection criteria based on the distribution of the shower shape, track--cluster matching, and consistency between the cluster energy and track momentum are then used in the identification of electron candidates~\cite{Khachatryan:2015hwa}. Muon candidates are reconstructed by requiring consistent hit patterns in the tracker and muon systems~\cite{Chatrchyan:2012xi}. Electron and muon candidates are required to be consistent with originating from the primary vertex by imposing restrictions on the size of their impact parameters in the transverse plane and longitudinal direction with respect to the beam axis. The relative isolation variable $I_\text{rel}$ for these candidates is defined as the scalar sum of the transverse momenta of all PF candidates, excluding the lepton, within a \pt-dependent cone size of radius $R$ around the candidate's trajectory, divided by the lepton \pt. The size $R$ depends on lepton \pt as follows:
\begin{equation}
R =
\begin{cases}
0.2, & \pt \leq 50\GeV, \\
10\GeV/\pt, & 50 < \pt < 200\GeV, \\
0.05, & \pt \geq 200\GeV.
\end{cases}
\end{equation}

The shrinking cone radius for higher-\pt leptons allows us to maintain high efficiency for the collimated decay products of boosted heavy objects. The isolation sum is corrected for contributions originating from pileup interactions through an area-based estimate~\cite{pileup} of the pileup energy deposited in the cone.

Hadronically decaying $\Pgt$ lepton ($\tauh$) candidates are reconstructed using the CMS hadron-plus-strips (HPS) algorithm~\cite{Khachatryan:2015dfa}. The constituents of the reconstructed jets are used to identify individual $\Pgt$ lepton decay modes with one charged hadron and up to two neutral pions, or three charged hadrons. The presence of
extra particles within the jet, not compatible with the reconstructed decay mode, is
used as a criterion to discriminate $\tauh$ decays from other jets.

Photon candidates are reconstructed from energy deposited in the ECAL, and selected using the distribution of the shower shape variable, the photon isolation, and the amount of leakage of the photon shower into the HCAL~\cite{Khachatryan:2015iwa}.

Monte Carlo (MC) simulations of events are used to study the properties of SM backgrounds and signal models. The \MADGRAPH5\_a\MCATNLO 2.2.2 generator~\cite{Alwall:2014hca} is used in leading-order (LO) mode to simulate events originating from \ttbar, \wjets, \zjets, \gjets, and quantum chromodynamics multijet processes~('QCD'), as well as signal events, based on LO NNPDF3.0~\cite{Ball:2014uwa} parton distribution functions (PDFs). Single top quark events produced in the \PQt\W~channel and \ttbar samples used in the single-lepton analysis are generated at next-to-leading order (NLO) with \textsc{Powheg} v2~\cite{Nason:2004rx,Frixione:2007vw,Alioli:2010xd,Re:2010bp}, while rare SM processes such as \ttbarZ~and \ttbarW~are generated at NLO using the \MADGRAPH5\_a\MCATNLO 2.2.2 program, using NLO NNPDF3.0 PDFs. Parton showering and hadronization is generated using \textsc{Pythia}8.205~\cite{Sjostrand:2014zea}. The response of the CMS detector for the SM backgrounds is simulated via the \GEANTfour~\cite{geant4} package. The CMS fast simulation package \cite{fastsim} is used to simulate all signal samples, and is verified to provide results that are consistent with those obtained from the full \GEANTfour-based simulation. Event reconstruction is performed in the same manner as for collision data. A nominal distribution of pileup interactions is used when producing the simulated samples. The samples are then reweighted to match the pileup profile observed in the collected data. The signal production cross sections are calculated using NLO with next-to-leading logarithm (NLL) soft-gluon resummation calculations~\cite{Borschensky:2014cia}. The most precise cross section calculations are used to normalize the SM simulated samples, corresponding most often to next-to-next-to-leading order (NNLO) accuracy.

\section{Search for top squarks in the fully-hadronic final state}
\label{sec:stop0l}
The top squark search in the all-jet final state is characterized by the categorization of events into exclusive search regions based on selection criteria applied to kinematic variables related to jets and \met, and the use of the CTT algorithm to identify boosted top quark candidates. The main backgrounds in the search regions are estimated from dedicated data control samples.

\subsection{Analysis strategy}
\label{sec:stop0l_strategy}
The events in this analysis are recorded using a trigger that requires the presence of two or more energetic jets within the tracker acceptance and large \met. To be efficient, events selected offline are therefore required to have at least two jets with $\pt > 75$\GeV, $\abs{\eta} < 2.4$, and $\met > 250$\GeV.
In order to reduce SM backgrounds with intrinsic \met such as leptonic $\ttbar$ and \wjets events, we reject events with isolated electrons or muons that have $\pt > 5$\GeV, $\abs{\eta} < 2.4$, and $I_\text{rel}$ less than 0.1 or 0.2, respectively. The contribution from events in which a \W~boson decays to a $\Pgt$ lepton is reduced by rejecting events containing isolated charged-hadron PF candidates with $\pt>10$\GeV and $\abs{\eta} < 2.5$ that are consistent with $\tauh$ decays. The isolation requirement applied is based on a discriminant obtained from a multivariate boosted decision tree (BDT) trained to distinguish the characteristics of charged hadrons originating from $\tauh$ decays. The transverse mass \MT of the system comprising the charged-hadron PF candidate and \ptvecmiss is required to be less than 100\GeV assuring consistency with $\tauh$ originating from a \W~boson decay, to minimize loss of signal at high jet multiplicity. The transverse mass for a particle q (in this case, the $\tauh$ candidate)  is defined as:
\begin{equation}
M_\mathrm{T}({q}, \ptvecmiss) \equiv \sqrt{2 q_{\mathrm{T}} \MET (1 - \cos \Delta\phi)},\label{eq:MT}
\end{equation}
with $q_{\mathrm{T}}$ denoting the particle transverse momentum, and $\Delta\phi$ the azimuthal separation between the particle and \ptvecmiss.

Events selected for the search sample must also have at least five jets with $\pt>20$\GeV, at least two of which must be \bq-tagged satisfying the loose working point of the CSV algorithm, with one or more of them required to additionally satisfy the medium working point. In addition, the absolute value of the azimuthal angle between \ptvecmiss and the closest of the four highest-\pt (leading) jets, $\dphijonetwothreefour$, must be at least 0.5.
 An imbalance in event \pt is produced in QCD events through a mismeasurement of jet $\pt$, and is often aligned with one of the leading jets in the event. The requirement on $\dphijonetwothreefour$ therefore greatly reduces the contribution of the QCD background. The set of selection criteria defined above will be referred to as the ``baseline selection'' for this search.

After imposing the baseline selection, we subdivide the event sample into categories based on kinematic observables related to jets and \met to improve the power of the analysis to discriminate between signal and the remaining SM background.
The dominant sources of SM background are \ttbar, \wjets, and \zjets events. The contribution from \ttbar and \wjets processes arises from events with \W~bosons decaying leptonically, in which the charged lepton either falls outside of the kinematic acceptance, or, in most cases, evades identification, and may be reconstructed as a jet. Large \met can be generated by the associated neutrino, allowing such events to satisfy the baseline selection criteria. This background is collectively referred to as the ``lost-lepton background''. Contributions arising from \ttbarW~and single top quark processes also enter this category, but with lesser importance. The contributions from \zjets and \ttbarZ~events arise when the \cPZ~boson decays to neutrinos, producing thereby a significant amount of \met. The QCD background is reduced to a subdominant level by the requirements of the baseline selection.

\begin{figure}[!htbp]
\centering
\includegraphics[width=0.48\textwidth]{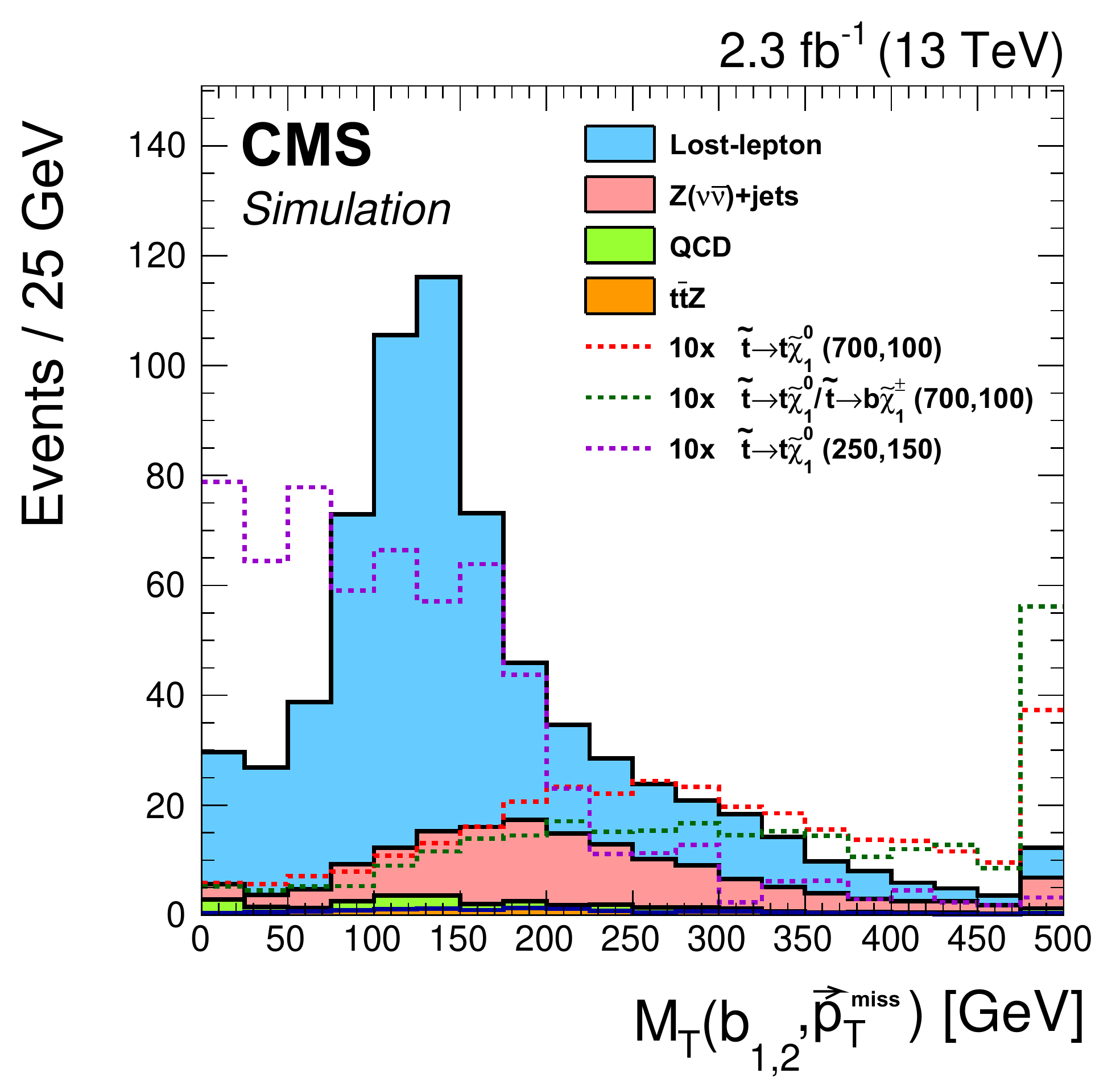}
\includegraphics[width=0.48\textwidth]{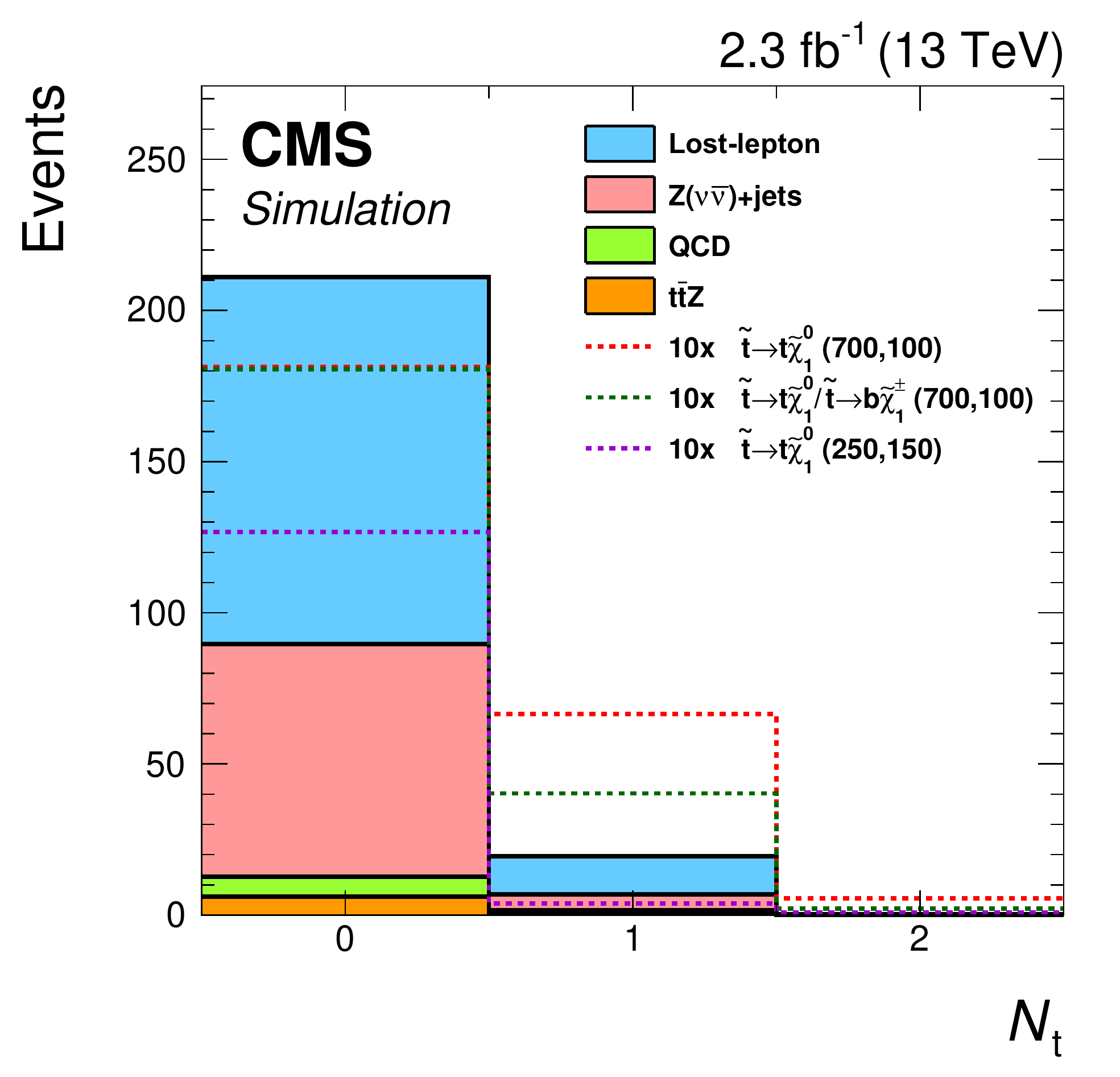}
\caption{\label{fig:catvars} The $\mtb$ distribution after the baseline selection of the top squark search in the all-jet final state (\cmsLeft), and the number of reconstructed top quarks for events in the high-$\mtb$ category (\cmsRight). Signal models with different top squark and neutralino mass hypotheses are shown, with the first number indicating the assumed top squark mass in units of\GeV and the second the neutralino mass. The expected signal yields are scaled up by a factor of 10 to facilitate comparison of the distributions with expectations from SM backgrounds. In this and subsequent figures, the last bin shown includes the overflow events.}
\end{figure}

In \ttbar events with a lost lepton, the transverse mass of the \bq~quark \ptvecmiss system arising from the same top quark decay as the lost lepton has a kinematic endpoint at the mass of the top quark. The observable~$\mtb$ is defined as
\begin{equation}
\mtb  \equiv  \min[ M_\mathrm{T}(\bq_{1}, \ptvecmiss), M_\mathrm{T}(\bq_{2}, \ptvecmiss) ],
\end{equation}
where $\bq_1, \bq_2$ are the two selected \bq-tagged jets with highest values in the CSV discriminant. Imposing a minimum requirement of 175\GeV on $\mtb$ reduces a significant portion of the \ttbar background, but also results in a loss in signal efficiency for models with small $\Delta m$, as seen in Fig.~\ref{fig:catvars}, in which signal models with different top squark and neutralino mass hypotheses are shown, with the first number indicating the assumed top squark mass in units of \GeV and the second the neutralino mass. To benefit from the separation power provided by this variable, we define two search categories, one with $\mtb\geq175$\GeV, taking advantage of the corresponding reduction in \ttbar background for signal models with large $\Delta m$, and another with $\mtb<175$\GeV to retain the statistical power of events populating the low-$\mtb$ region for models with small $\Delta m$.

Signal events with all-jet top quark decays should have at least six jets in the final state, although in the case of signals with compressed mass spectra these jets can be too soft in \pt to satisfy the jet selection threshold. Additional jets may be produced through initial-state radiation (ISR). The jet multiplicity is lower for the semileptonic \ttbar background, as well as for the other backgrounds remaining after the baseline selection. A requirement of higher reconstructed jet multiplicity therefore improves the discrimination of signal events from the SM background. We consider two regions in jet multiplicity for the analysis, a high-$\nj$~region ($\geq 7$ jets) that benefits from this improved discrimination, and a medium-$\nj$~region (5--6 jets) to preserve signal events with fewer reconstructed jets. The high-$\nj$~region in conjunction with the low threshold on the $\pt$~of selected jets improves sensitivity for signal models with soft decay products in the final state.

In the high-$\mtb$ category, requiring the presence of at least one top quark reconstructed by the CTT algorithm ($\nt \geq 1$) ensures a high-purity selection of signal events with highly boosted top quarks, at the sacrifice of some loss in signal efficiency. To benefit from this high-purity region, without giving up signal events that would enter the $\nt = 0$ region, we use both regions to extract the final signal. Figure~\ref{fig:catvars} shows the $\nt$ distribution for events in the high-$\mtb$ category. Subdividing each $\nt$ region by the number of \bq-tagged jets ($\nb$) that satisfy the medium working point of the CSV algorithm provides even greater discrimination of signal from background.
Since there are relatively few events in the $\nt \geq 1$ category, the subcategorization in $\nj$~is not performed for these events because it provides no additional gain after the $\nb$~subdivision.

The event categorization according to $\mtb$, $\nj$, $\nb$, and $\nt$ is summarized in Table~\ref{tab:evtcategories}. In each of these categories, we use \met as the final discriminant to characterize and distinguish potential signal from the SM background by defining five \met regions.
The analysis is therefore carried out in a total of 50 disjoint search regions (SRs).

\begin{table}[htb]
\centering
\topcaption{\label{tab:evtcategories} Categorization in $\mtb$, $\nj$, $\nb$, and $\nt$ used to define the SRs for the top squark search in the all-jet final state. Events in each category are further separated into the following \met regions: 250--300, 300--400, 400--500, 500--600, and $>$600\GeV, resulting in 50 disjoint SRs.}
{\begin{tabular}{c|cc|c}
\hline
\multicolumn{2}{c}{$\mtb<175$\GeV} & \multicolumn{2}{c}{$\mtb\geq175$\GeV}\\
\hline
$\nb = 1$ & $\nb \geq 2$ & $\nb = 1$ & $\nb \geq 2$ \\
\hline
\multirow{3}{*}{$5 \leq \nj \leq 6$} & \multirow{3}{*}{$5 \leq \nj \leq 6$} & \multicolumn{2}{c}{$\nt = 0$} \\
\cline{3-4}
& & $5 \leq \nj \leq 6$ & $5 \leq \nj \leq 6$ \\
 & & $\nj \geq 7$ & $\nj \geq 7$ \\
\cline{3-4}
\multirow{2}{*}{$\nj \geq 7$} & \multirow{2}{*}{$\nj \geq 7$} & \multicolumn{2}{c}{$\nt \geq 1$}\\
\cline{3-4}
& & $\nj \geq 5$ & $\nj \geq 5$ \\
\hline
\end{tabular}
}

\end{table}

\subsection{Background estimation}
\label{sec:stop0l_bkgest}

\subsubsection{Estimation of the lost-lepton background}
\label{sec:stop0l_llbkg}
The lost-lepton background is estimated from a single-lepton control sample, selected using the same trigger as the search sample, and consisting of events that have at least one lepton~($\ell$) obtained by inverting the electron and muon rejection criteria. Studies in simulation indicate that the event kinematics for different lepton flavours are similar enough to estimate them collectively from the same control sample. Potential signal contamination is suppressed by requiring $\mtl<100$\GeV. If there is more than one lepton satisfying the selection criteria, the lepton used to determine $\mtl$ is chosen randomly.  The events selected in the lepton control sample are further subdivided into control regions (CRs) using the same selection criteria as in the search sample, according to $\mtb$, $\nj$, $\nt$, and $\met$. However with the requirement $\nb\geq1$ the distribution in \met originating from lost-lepton processes is independent of $\nb$, and therefore the CRs are not subdivided according to the number of \bq-tagged jets. These CRs generally have a factor of 2--4 more events than the corresponding SRs.

The estimation of the lost-lepton background in each SR is based on the event count in data in the corresponding single-lepton CR ($N^{\text{data}}_{1\ell}$). We translate this event count to the SR by means of a lost-lepton transfer factor $\tfllb$ obtained from simulation. The lost-lepton background prediction can therefore be extracted as
\begin{equation}
N^\text{pred}_{\text{LL}} = N^{\text{data}}_{1\ell}~\tfllb,\quad \tfllb = \frac{N^{\text{sim}}_{0\ell}}{N^{\text{sim}}_{1\ell}},
\end{equation}
where $N^{\text{sim}}_{0\ell}$ and $N^{\text{sim}}_{1\ell}$ are the simulated lost-lepton background yields in the corresponding zero- and single-lepton regions, respectively, taking into account contributions from \ttbar and \wjets events, with smaller contributions from single top quark and \ttbarW~processes. The contamination from other SM processes in the single-lepton CRs is found to be negligible in studies of simulated events. Monte Carlo simulated samples are used to estimate the small component of the lost-lepton background that originates from leptons falling outside the kinematic acceptance, since this component is not accounted for in the CRs.

To improve the statistical power of the estimation, CRs with $\nt\geq1$ are summed over \met bins as well as over $\nb$. We rely on the simulation through $N^\text{sim}_{0\ell}$ to provide the \met-dependence and to predict the yield in each of the SRs with $\nt\geq1$.  We check this procedure by computing the data-to-simulation ratios $N^\text{data}_{1\ell}/N^\text{sim}_{1\ell}$ in the higher-statistics region of $\mtb\geq175$\GeV with $\nt = 0$, and find no evidence of a dependence on \met. We assign the relative statistical uncertainties of these ratios as systematic uncertainties in the SRs.

The dominant uncertainty in the lost-lepton prediction is due to the limited number of events in the CRs, and can be as large as $100\%$. The statistical uncertainties in the simulated samples also affect the uncertainty in the prediction via the transfer factors. The effect in the uncertainty ranges between $3\%$ and $50\%$. A source of bias in the prediction can arise from a possible difference between data and simulation in the background composition, which is assessed by independently changing the cross sections of the \wjets and $\ttbar$ processes by ${\pm}20\%$ based on CMS differential cross section measurements~\cite{Khachatryan:2014uva,CMS-PAS-TOP-16-008}. The effect of these changes is as large as $11\%$ for the transfer factors. The uncertainties in the measurements of correction factors in lepton efficiency that are applied to the simulation to reduce discrepancies with the data lead to a systematic uncertainty of up to $7\%$ in $\tfllb$. All other sources of systematic uncertainty, to be discussed in Section~\ref{sec:systematics}, have a negligible effect on the prediction.

\subsubsection{Estimation of the \texorpdfstring{$\znunu$}{Znunu} background}
\label{sec:stop0l_znunu}
Two methods are traditionally used to estimate the \znunu~background in searches involving all-jet final states with large \met. The first method relies on a sample dominated by \zll+jets events, which has the advantage of accessing very similar kinematics to the \znunu~process, after correcting for the difference in acceptance between charged-lepton pairs and pairs of neutrinos, but is statistically limited in regions defined with stringent requirements on jets and \met. The second method utilizes \gjets events that have a significantly larger production cross section than the \zll+jets process, but similar leading-order Feynman diagrams. The two main differences between the processes that must be taken into account, namely, different quark-boson couplings and the massive nature of the \cPZ~boson, become less important at large \cPZ~boson \pt, which is the kinematic region we are probing in this search.

We have therefore adopted  a hybrid method to estimate the \znunu~background by combining information from \zjets, with \zll, and \gjets events. \zll~events are used to obtain the normalization for the $\znunu$ background in different ranges of $\nb$ to account for potential effects related to heavy-flavour production, while the much higher yields from the \gjets sample are exploited to extract corrections to distributions of variables used to characterize the SRs. The \zll~events are obtained from dielectron and dimuon triggers, with the leading lepton required to have $\pt>20$\GeV, and the trailing lepton $\pt>15$ and $>10$\GeV for electrons and muons, respectively.  Both leptons must also have  $\abs{\eta} < 2.4$. The \gjets sample is collected through a single-photon trigger, and consists of events containing photons with $\pt>180$\GeV and $\abs{\eta} < 2.5$. The transverse momentum of the dilepton or photon system is added vectorially to \ptvecmiss in each event of the corresponding data samples to emulate the kinematics of the \znunu~process. The modified \met, denoted by \metll and \metg for  the \zll~and \gjets processes, respectively, is used to calculate related kinematic variables.

The prediction for the \znunu~background is given by:
\begin{equation}
N^{\text{pred}}_{\znunuM} = N^{\text{sim}}_{\znunuM} R_{\cPZ} S_{\gamma},
\end{equation}
where $N^{\text{sim}}_{\znunuM}$ is the expected number of \znunu~events obtained from simulation, $R_{\cPZ}$ is the flavour-dependent \zjets normalization factor measured with the \zll~~sample, and $S_{\gamma}$ is the  correction factor for distributions in \met and jet kinematic variables extracted from the \gjets sample.
The underlying assumption of this hybrid estimation method is that the differences in the \met (or \metg) distributions between data and simulation are similar for $\znunu$ and photon events. We checked this assumption by comparing the ratios of data to simulation observed in the \metll and \metg distributions for \zll+jets and \gjets samples, respectively, and found them to agree.

The factor $R_{\cPZ}$ is calculated by comparing the observed and expected \zll~yields for a relaxed version of the baseline selection. In particular, we remove the requirements on $\dphijonetwothreefour$ after confirming that this does not bias the result, and relax the requirements on \metll from a threshold of 250\GeV to a threshold of 100\GeV. To increase the purity of the \zll~events, we require the dilepton invariant mass to lie within the \cPZ~boson mass window of $80<M_{\ell\ell}<100$\GeV. The normalization of the nonnegligible \ttbar~contamination is estimated in the region outside the \cPZ~boson mass window ($20<M_{\ell\ell}<80$ or $M_{\ell\ell}>100$\GeV) and taken into account. Small contributions from $\PQt\cPZ$ and \ttbarZ~production, estimated from simulation, are included in the \zll~sample when measuring $R_{\cPZ}$. Contributions from $\PQt\W$ and \ttbarW~are included in the simulation sample used to obtain the normalization factor for the \ttbar contamination. As discussed previously, we calculate $R_{\cPZ}$ separately for different $\nb$ requirements. The values obtained are $0.94\pm0.13$ and $0.84\pm0.19$ for $\nb = 1$ and ${\geq}2$, respectively. The uncertainty in $R_{\cPZ}$  originates from the limited event counts in data and simulation, and from the extrapolation in \met.

The quantity $S_{\gamma}$ is the correction factor related to the modelling of the distributions in the kinematic variables of \znunu~events. It is calculated via a comparison of the \metg distributions of \gjets events in simulation and data. The simulation is normalized to the number of events in data after applying the baseline selection. To suppress potential contamination from signal and avoid overlap with the search sample, we only consider events with $\met<200\GeV$. The $S_{\gamma}$ factor is estimated separately for each SR to account for any potential mismodelling of the observables $M_{\mathrm{T}}(\PQb_{1,2},\metg)$, $\nj$, $\metg$, and $\nt$ in simulation. Since no statistically significant dependence of \metg on $\nb$ is observed, we improve the statistical power of the correction by combining the $\nb=1$ and $\nb \geq 2$ subsets of the \gjets sample to extract the $S_{\gamma}$ corrections. The correction factors range between 0.3 and 2, with uncertainties of up to 100\% due to the limited number of events in the data sample.

The \gjets control data have contributions from three main components: prompt photons produced directly or via fragmentation, and other objects misidentified as photons. The prompt photon purity measured in Ref.~\cite{CMS-PAS-SUS-15-003} shows good agreement between data and simulation. In addition, the impact of varying the fraction of misidentified photons, or those produced via fragmentation, by $50\%$ in simulated events results in a bias of less than $5\%$ in the \met distribution from the predicted \znunu~background. We therefore rely on simulation to estimate the relative contributions of the three different components.

The statistical uncertainty in the \gjets control data and the uncertainty in $R_{\cPZ}$ are the main sources of uncertainty in the \znunu~prediction. The statistical uncertainties in the simulated samples, ranging up to $50\%$ in both the SRs and in the \gjets CRs, also makes sizeable contributions.

\subsubsection{Estimation of the QCD background}
\label{sec:stop0l_qcd}
The QCD background is estimated using a data CR selected with the same trigger as the SR and enriched in QCD events by imposing a threshold on the azimuthal separation between \ptvecmiss and the closest of the three leading jets, namely $\dphijonetwothree < 0.1$. After correcting for the contribution from other SM processes (\ie \ttbar and \wjets), estimated by applying the normalization factor obtained in the corresponding single-lepton control sample to simulation, we translate the observation in this CR to a prediction in the SR by means of transfer factors obtained from simulation. Each transfer factor is defined as the ratio of the expected QCD events satisfying $\dphijonetwothreefour > 0.5$ to the expected QCD events with $\dphijonetwothree < 0.1$. The estimation is carried out in each search category. Since the distributions in key observables show little dependence on $\nb$, the QCD CR is summed over $\nb$ to improve the statistical precision of the estimation.

The main source of QCD events populating the SR is from severe mismeasurement of the \pt~of one or more jets in the event. Correct modelling of jet mismeasurement in simulation is therefore an important part of the QCD prediction. The level of mismeasurement of a simulated event is parameterized by the jet response of the most mismeasured jet, which is the jet with the greatest absolute difference between the reconstructed and generated $\pt$. The jet response, $\rjet$, is defined as the ratio of the reconstructed $\pt$ of a jet to its generated \pt, computed without including the loss of visible momentum due to neutrinos. We use the observable $\prjet$, defined as the ratio of the $\pt$ of a jet to the magnitude of the vector sum of its transverse momentum and $\ptvecmiss$, as an approximate measure of the true jet response in data, and extract mismeasurement correction factors for the simulation by comparing $\prjet$ of the jet closest in $\phi$ to $\ptvecmiss$ between data and simulation. The correction factors extracted from simulation are parameterized by $\rjet$ and the flavour of the most mismeasured jet. The correction factors range between 0.44 and 1.13, and are applied in the simulation on an event-by-event basis.

The largest sources of uncertainty in the QCD prediction originate from the limited event counts in data and simulated samples surviving the selection, giving rise to uncertainties of up to 100\% in the estimated QCD background contribution in some SRs. The uncertainty due to jet response corrections is up to 15\%, while the uncertainty due to contributions from non-QCD processes in the data CR ranges from 7\% to 35\%.

\subsubsection{Estimation of the \texorpdfstring{$\ttbarZ$}{ttbar Z}  background}
\label{sec:stop0l_ttz}
Contributions from the $\ttbarZ$ process are generally small since this is a relatively rare process. However, it has a final state very similar to signal when the \cPZ~boson decays to neutrinos and both top quarks decay only into jets, which can constitute up to 25\% of the total SM background in some SRs with large $\met$ and $\nt \geq 1$. The $\ttbarZ$ prediction is obtained from simulation. We assign a 30$\%$ uncertainty to the \ttbarZ~cross section, based on the $8\TeV$ CMS measurement~\cite{Khachatryan:2015sha}. Additional theoretical and experimental uncertainties in the prediction are evaluated as will be discussed in Section~\ref{sec:systematics}, and range up to 25\% and 20\%, respectively, depending on the SR. We also take into consideration the statistical uncertainty in the simulation, which ranges from 5\% to 100\% for regions with small $\ttbarZ$ contributions.

\subsection{Results}
\label{sec:stop0l_results}
\begin{figure*}[hp]
\centering
\includegraphics[width=0.49\textwidth]{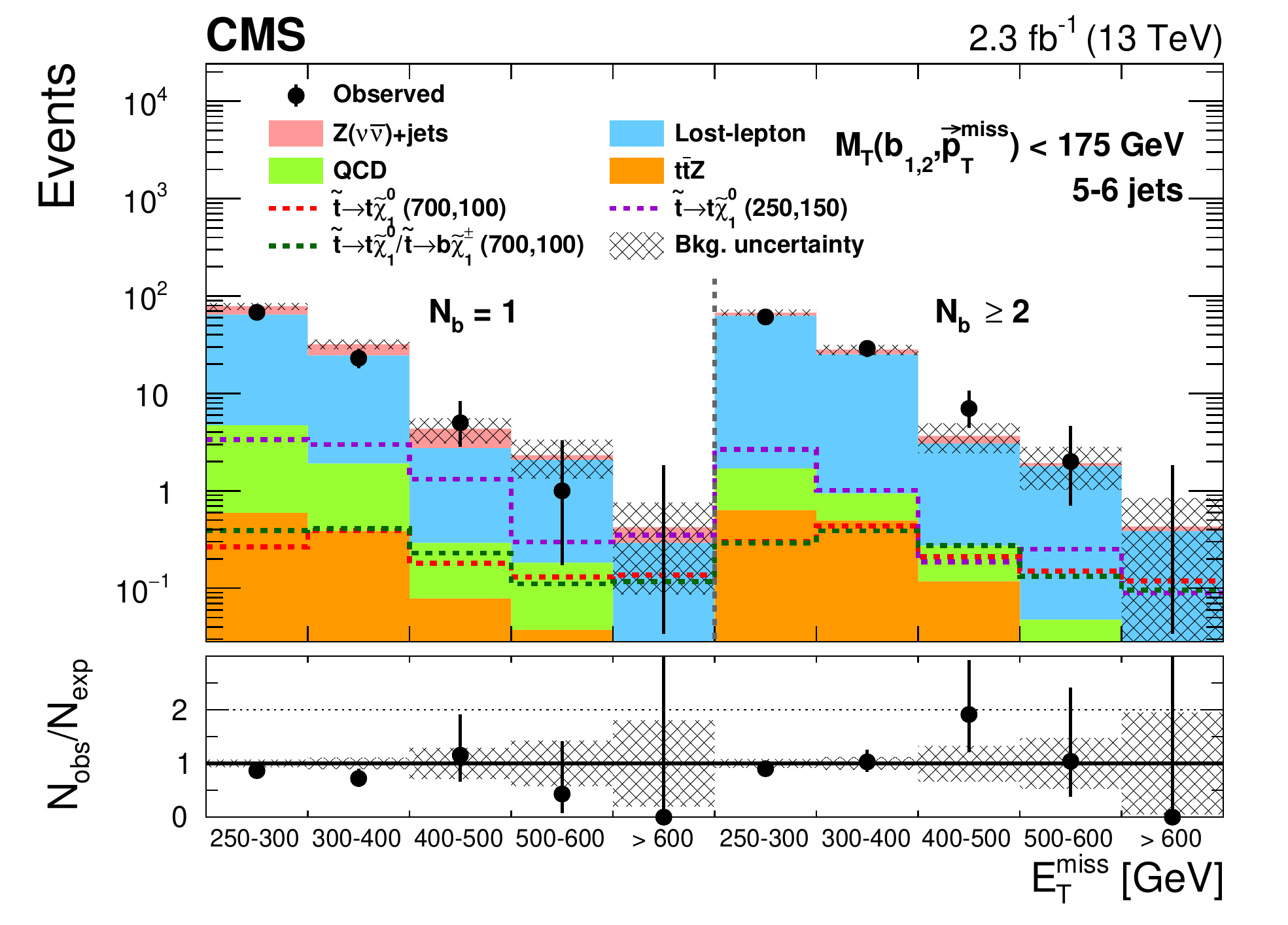}
\includegraphics[width=0.49\textwidth]{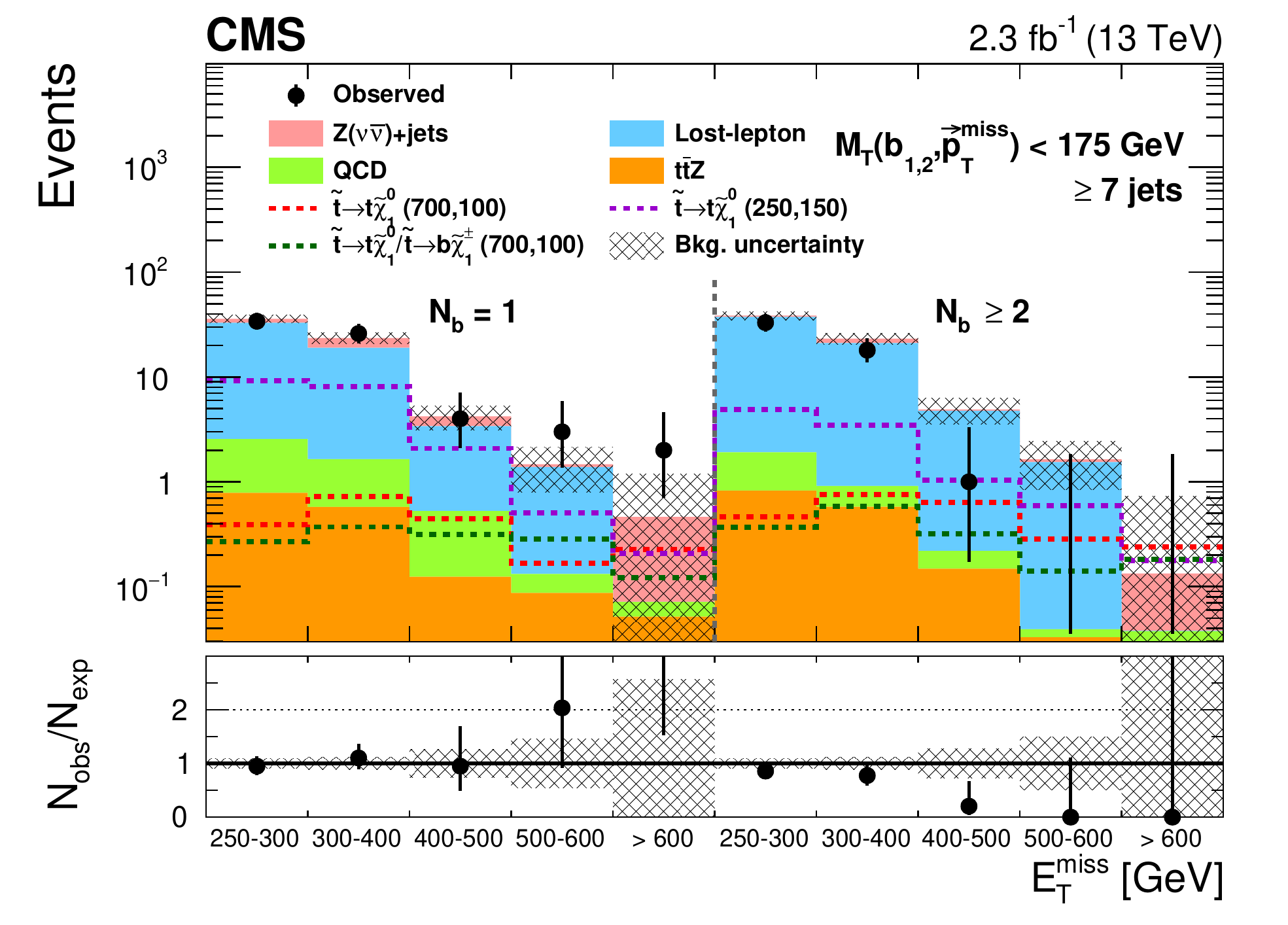}
\includegraphics[width=0.49\textwidth]{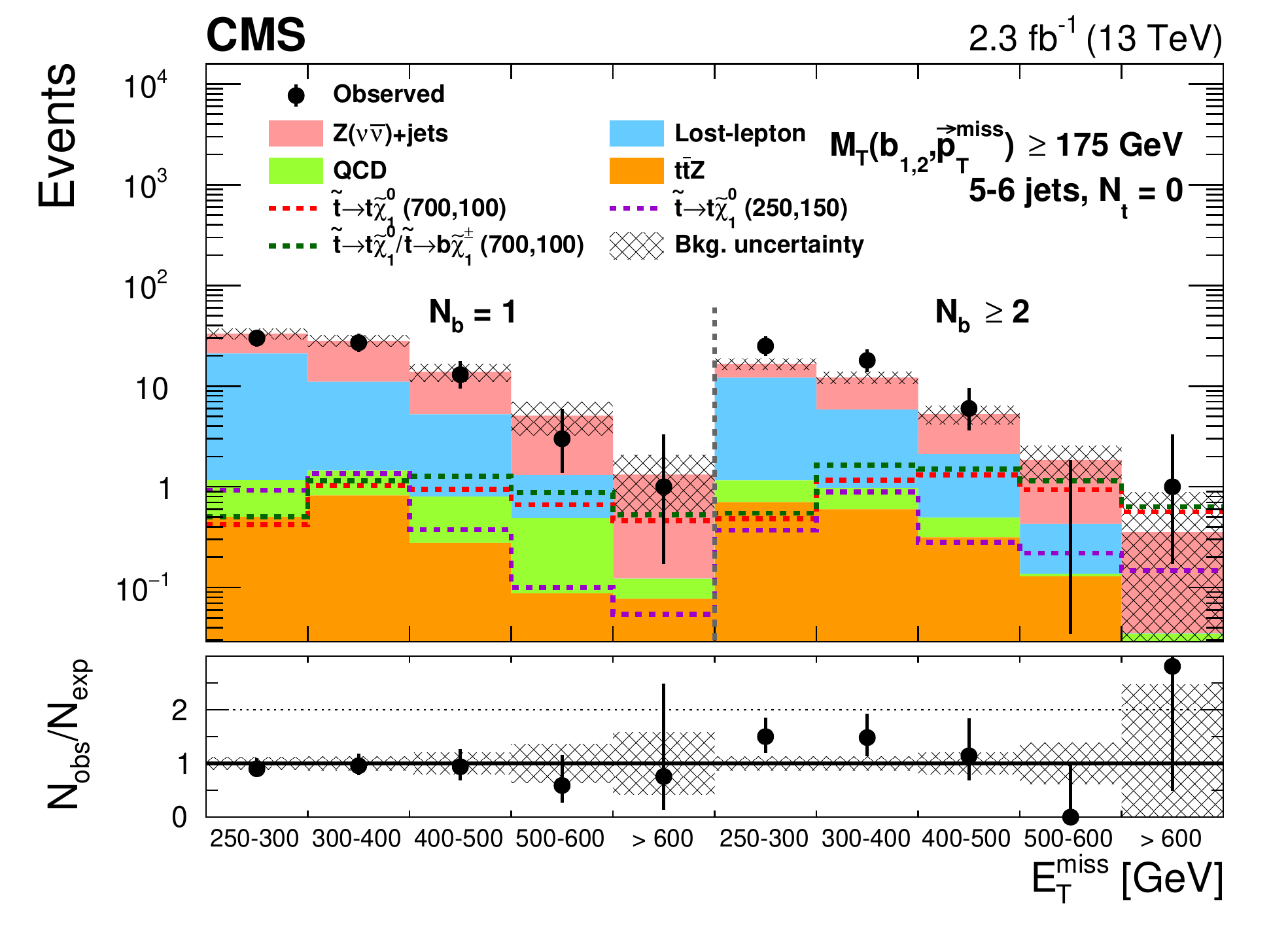}
\includegraphics[width=0.49\textwidth]{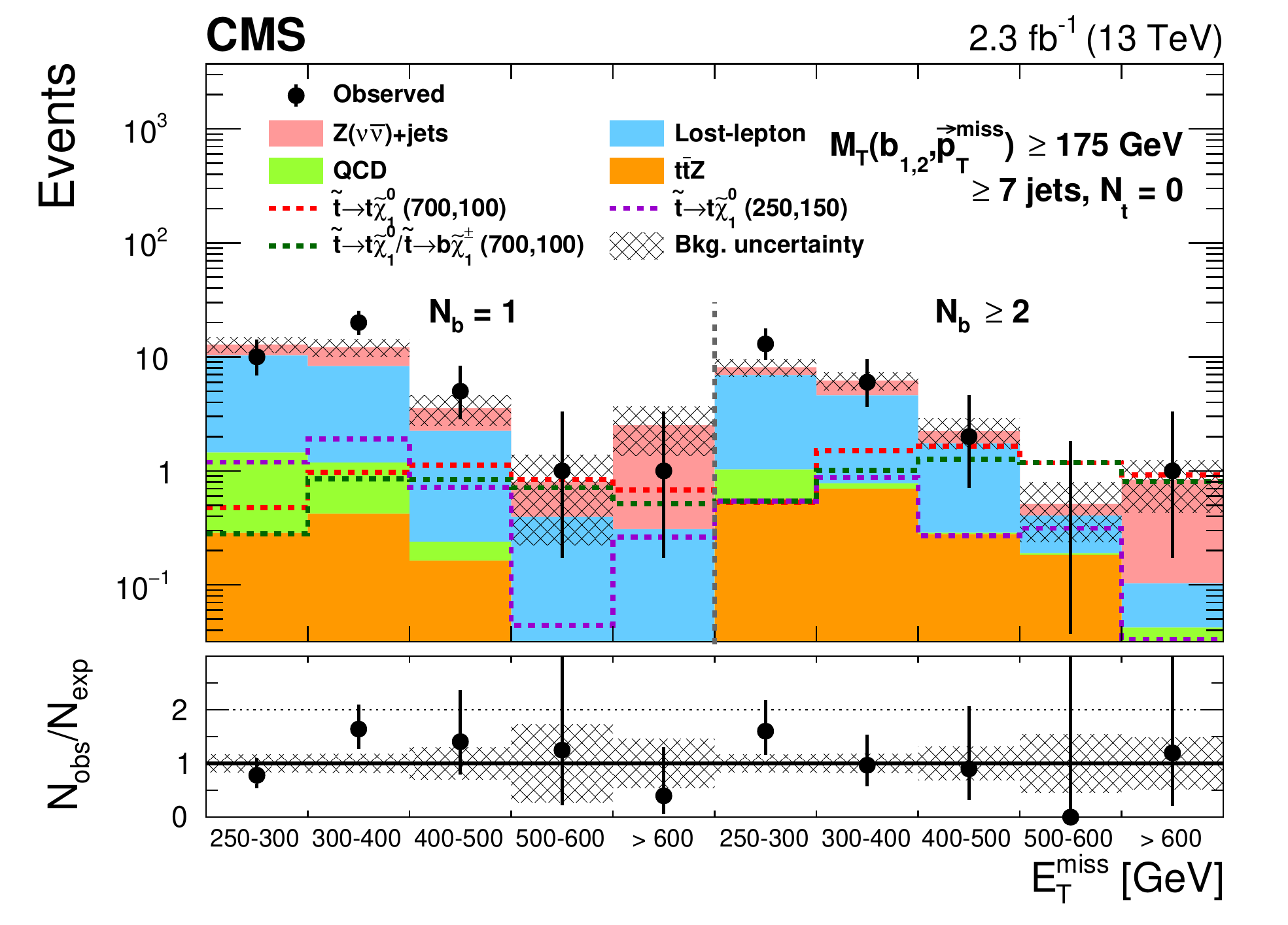}
\includegraphics[width=0.49\textwidth]{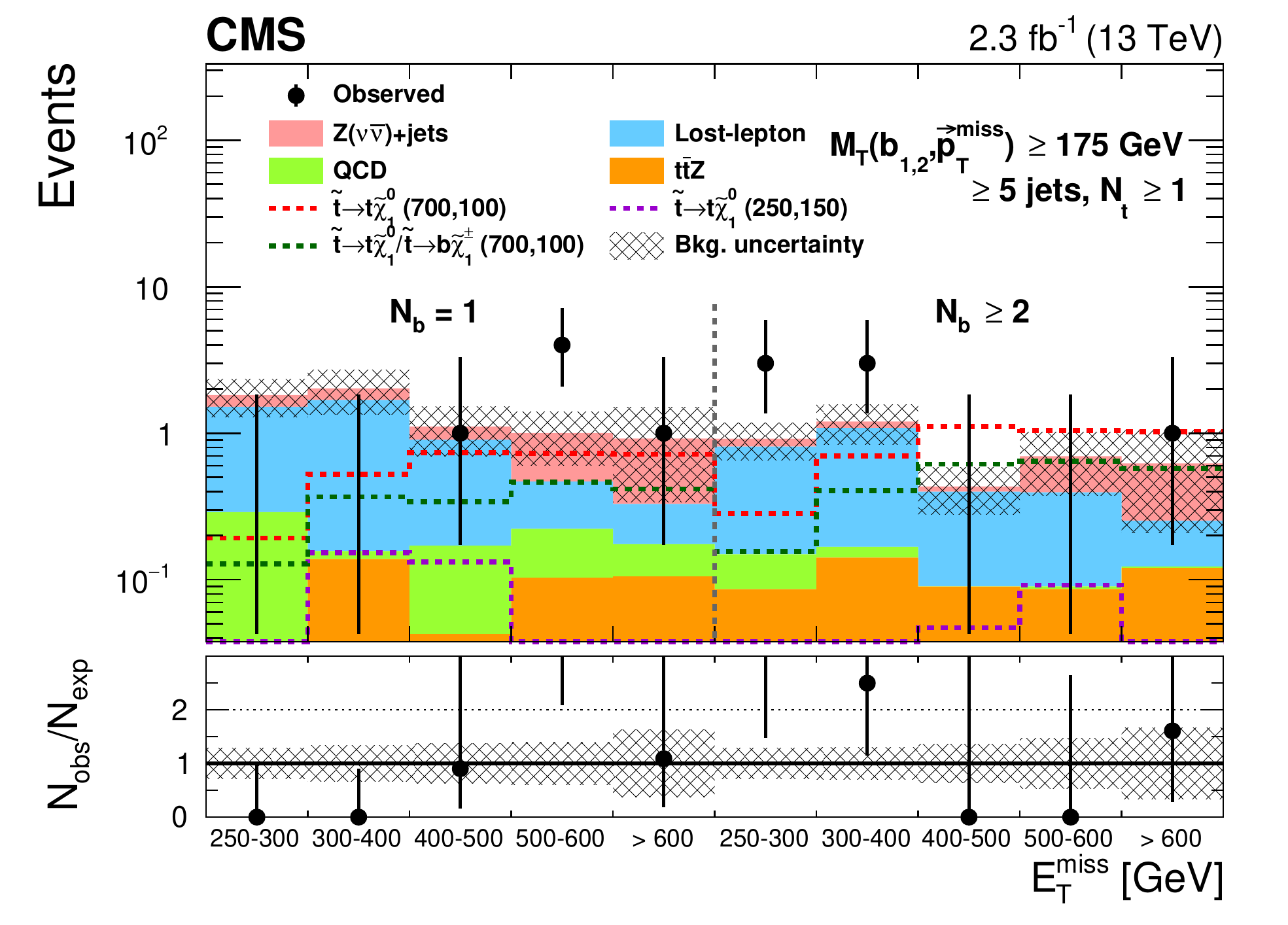}
\topcaption{Observed and estimated SM background and signal yields in the SRs of the top squark search in the all-jet final state: $\mtb<175$\GeV, $5 \leq \nj \leq 6$ (upper left), $\mtb<175$\GeV, $\nj \geq 7$ (upper right), $\mtb\geq175$\GeV, $\nt = 0, 5 \leq \nj \leq 6$ (middle left),  $\mtb\geq175$\GeV, $\nt = 0, \nj \geq 7$ (middle right), $\mtb\geq175$\GeV, $\nt \geq 1, \nj \geq 5$ (bottom row). The first 5 bins in each plot correspond to $\met$ ranges of 250--300, 300--400, 400--500, 500--600, $>600$\GeV for $\nb = 1$, and the second 5 bins correspond to the same \met binning for $\nb \geq 2$. The SM background predictions shown do not include the effects of the maximum likelihood fit to the data. The ratio of the data to the SM prediction extracted from CRs is shown in the lower panel of each plot. The shaded black band represents the statistical and systematic uncertainty in the background prediction.
}
\label{fig:sryields}
\end{figure*}

Figure~\ref{fig:sryields} shows the yields in each of the SR bins, as well as the predicted SM backgrounds based on the background estimation methods discussed in Section~\ref{sec:stop0l_bkgest}.  The results are also summarized in Table~\ref{tab:all-bin-yields}. Expected yields are also shown for two benchmark  models for the pure $\stopq \to \topq^{(*)} \lsp$ decay and one for the mixed ($\stopq\to \topq \lsp$ or $\stopq\to \bq \chipmone$) decay. No statistically significant deviation from the SM prediction is observed in the data.

\begin{table*}[htbp]
\centering
\topcaption{\label{tab:all-bin-yields} Observed and predicted background yields in the different search regions for the top squark search in the all-jet final state.  The total uncertainty is given for each background prediction. }
\cmsTable{
\begin{tabular}{ c  cccc  c  c }
\hline
\met [\GeVns{}]  &  Lost-lepton  &  \znunu  &  QCD  &  \ttbarZ  &  Total SM  &  Data  \\
\hline
\multicolumn{7}{c}{$\mtb < 175$\GeV, $5\leq\nj\leq6$, $\nb = 1$} \\
\hline
250-300 & 60 $\pm$ 6 & 14 $\pm$ 3 & 4.1 $\pm$ 1.7 & 0.59 $\pm$ 0.21 & 79 $\pm$ 7 & 68 \\
300-400 & 23 $\pm$ 3 & 7.4 $\pm$ 1.9 & 1.5 $\pm$ 0.8 & 0.39 $\pm$ 0.14 & 32 $\pm$ 4 & 23 \\
400-500 & 2.5 $\pm$ 1.0 & 1.6 $\pm$ 0.8 & 0.21 $\pm$ 0.15 & 0.08 $\pm$ 0.04 & 4.3 $\pm$ 1.3 & 5 \\
500-600 & 1.9 $\pm$ 1.0 & 0.25 $^{+0.27}_{-0.25}$ & 0.14 $^{+0.15}_{-0.14}$ & 0.04 $\pm$ 0.02 & 2.3 $\pm$ 1.0 & 1 \\
$>$600 & 0.28 $^{+0.31}_{-0.28}$ & 0.13 $^{+0.15}_{-0.13}$ & 0.01 $\pm$ 0.01 & $<$0.01 & 0.42 $\pm$ 0.34 & 0 \\
\hline
\multicolumn{7}{c}{$\mtb < 175$\GeV, $5\leq\nj\leq6$, $\nb \geq 2$} \\
\hline
250-300 & 61 $\pm$ 6 & 4.7 $\pm$ 1.4 & 1.1 $\pm$ 0.5 & 0.63 $\pm$ 0.22 & 68 $\pm$ 6 & 61 \\
300-400 & 24 $\pm$ 3 & 3.0 $\pm$ 1.0 & 0.44 $\pm$ 0.23 & 0.50 $\pm$ 0.18 & 28 $\pm$ 4 & 29 \\
400-500 & 2.8 $\pm$ 1.2 & 0.61 $\pm$ 0.33 & 0.16 $\pm$ 0.13 & 0.12 $\pm$ 0.06 & 3.7 $\pm$ 1.2 & 7 \\
500-600 & 1.7 $\pm$ 0.9 & 0.13 $^{+0.15}_{-0.13}$ & 0.05 $^{+0.06}_{-0.05}$ & $<$0.01 & 1.9 $\pm$ 0.9 & 2 \\
$>$600 & 0.38 $^{+0.41}_{-0.38}$ & 0.04 $^{+0.06}_{-0.04}$ & $<$0.01 & 0.01 $\pm$ 0.01 & 0.43 $\pm$ 0.41 & 0 \\
\hline
\multicolumn{7}{c}{$\mtb < 175$\GeV, $\nj\geq7$, $\nb = 1$} \\
\hline
250-300 & 30 $\pm$ 4 & 3.0 $\pm$ 1.0 & 1.8 $\pm$ 0.6 & 0.79 $\pm$ 0.28 & 36 $\pm$ 4 & 34 \\
300-400 & 17 $\pm$ 3 & 4.6 $\pm$ 1.6 & 1.1 $\pm$ 0.5 & 0.58 $\pm$ 0.21 & 24 $\pm$ 3 & 26 \\
400-500 & 2.9 $\pm$ 0.9 & 0.82 $\pm$ 0.64 & 0.40 $\pm$ 0.27 & 0.12 $\pm$ 0.07 & 4.2 $\pm$ 1.1 & 4 \\
500-600 & 1.3 $\pm$ 0.7 & 0.09 $^{+0.11}_{-0.09}$ & 0.05 $\pm$ 0.05 & 0.09 $\pm$ 0.05 & 1.5 $\pm$ 0.7 & 3 \\
$>$600 & $<$0.56 & 0.39 $^{+0.46}_{-0.39}$ & 0.02 $\pm$ 0.02 & 0.05 $\pm$ 0.03 & 0.46 $^{+0.72}_{-0.46}$ & 2 \\
\hline
\multicolumn{7}{c}{$\mtb < 175$\GeV, $\nj\geq7$, $\nb \geq 2$} \\
\hline
250-300 & 36 $\pm$ 4 & 0.96 $\pm$ 0.38 & 1.1 $\pm$ 0.5 & 0.83 $\pm$ 0.30 & 38 $\pm$ 4 & 33 \\
300-400 & 20 $\pm$ 3 & 2.1 $\pm$ 0.9 & 0.34 $\pm$ 0.19 & 0.58 $\pm$ 0.22 & 23 $\pm$ 3 & 18 \\
400-500 & 4.5 $\pm$ 1.4 & 0.15 $\pm$ 0.13 & 0.07 $\pm$ 0.05 & 0.15 $\pm$ 0.07 & 4.9 $\pm$ 1.4 & 1 \\
500-600 & 1.5 $\pm$ 0.8 & 0.09 $^{+0.11}_{-0.09}$ & 0.01 $\pm$ 0.01 & 0.03 $\pm$ 0.03 & 1.6 $\pm$ 0.8 & 0 \\
$>$600 & $<$0.59 & 0.10 $^{+0.12}_{-0.10}$ & 0.01 $\pm$ 0.01 & 0.03 $\pm$ 0.02 & 0.13 $^{+0.60}_{-0.13}$ & 0 \\
\hline
\multicolumn{7}{c}{$\mtb \geq 175$\GeV, $5\leq\nj\leq6$, $N_\mathrm{t} = 0$, $\nb = 1$} \\
\hline
250-300 & 20 $\pm$ 3 & 12 $\pm$ 3 & 0.66 $\pm$ 0.37 & 0.50 $\pm$ 0.19 & 33 $\pm$ 5 & 30 \\
300-400 & 9.6 $\pm$ 2.3 & 17 $\pm$ 4 & 0.63 $\pm$ 0.32 & 0.82 $\pm$ 0.27 & 28 $\pm$ 4 & 27 \\
400-500 & 4.4 $\pm$ 1.9 & 8.6 $\pm$ 2.6 & 0.52 $\pm$ 0.35 & 0.28 $\pm$ 0.12 & 14 $\pm$ 3 & 13 \\
500-600 & 0.82 $\pm$ 0.63 & 3.8 $\pm$ 1.8 & 0.40 $\pm$ 0.35 & 0.09 $\pm$ 0.06 & 5.1 $\pm$ 1.9 & 3 \\
$>$600 & $<$0.4 & 1.2 $\pm$ 0.7 & 0.05 $\pm$ 0.05 & 0.08 $\pm$ 0.04 & 1.3 $\pm$ 0.8 & 1 \\
\hline
\multicolumn{7}{c}{$\mtb \geq 175$\GeV, $5\leq\nj\leq6$, $N_\mathrm{t} = 0$, $\nb \geq 2$} \\
\hline
250-300 & 11 $\pm$ 2 & 4.5 $\pm$ 1.4 & 0.45 $\pm$ 0.27 & 0.70 $\pm$ 0.24 & 17 $\pm$ 3 & 25 \\
300-400 & 4.9 $\pm$ 1.2 & 6.3 $\pm$ 1.8 & 0.37 $\pm$ 0.23 & 0.60 $\pm$ 0.22 & 12 $\pm$ 2 & 18 \\
400-500 & 1.6 $\pm$ 0.7 & 3.1 $\pm$ 1.1 & 0.18 $\pm$ 0.17 & 0.31 $\pm$ 0.12 & 5.3 $\pm$ 1.4 & 6 \\
500-600 & 0.29 $\pm$ 0.24 & 1.4 $\pm$ 0.8 & 0.01 $\pm$ 0.01 & 0.13 $\pm$ 0.06 & 1.9 $\pm$ 0.8 & 0 \\
$>$600 & $<$0.49 & 0.32 $\pm$ 0.20 & 0.01 $^{+0.02}_{-0.01}$ & 0.02 $\pm$ 0.02 & 0.36 $^{+0.53}_{-0.36}$ & 1 \\
\hline
\multicolumn{7}{c}{$\mtb \geq 175$\GeV, $\nj\geq7$ $N_\mathrm{t} = 0$, $\nb = 1$} \\
\hline
250-300 & 8.8 $\pm$ 1.9 & 2.5 $\pm$ 1.0 & 1.2 $\pm$ 0.6 & 0.29 $\pm$ 0.18 & 13 $\pm$ 2 & 10 \\
300-400 & 7.1 $\pm$ 1.8 & 3.9 $\pm$ 1.5 & 0.76 $\pm$ 0.46 & 0.42 $\pm$ 0.18 & 12 $\pm$ 2 & 20 \\
400-500 & 2.0 $\pm$ 0.8 & 1.3 $\pm$ 0.7 & 0.08 $\pm$ 0.07 & 0.16 $\pm$ 0.09 & 3.6 $\pm$ 1.1 & 5 \\
500-600 & 0.38 $^{+0.40}_{-0.38}$ & 0.40 $^{+0.43}_{-0.40}$ & 0.02 $\pm$ 0.02 & $<$0.01 & 0.80 $\pm$ 0.59 & 1 \\
$>$600 & 0.28 $^{+0.33}_{-0.28}$ & 2.2 $\pm$ 1.2 & 0.02 $^{+0.03}_{-0.02}$ & $<$0.01 & 2.5 $\pm$ 1.2 & 1 \\
\hline
\multicolumn{7}{c}{$\mtb \geq 175$\GeV, $\nj\geq7$ $N_\mathrm{t} = 0$, $\nb \geq 2$} \\
\hline
250-300 & 5.9 $\pm$ 1.3 & 1.2 $\pm$ 0.5 & 0.46 $\pm$ 0.24 & 0.57 $\pm$ 0.21 & 8.1 $\pm$ 1.5 & 13 \\
300-400 & 3.8 $\pm$ 1.0 & 1.6 $\pm$ 0.7 & 0.08 $\pm$ 0.06 & 0.70 $\pm$ 0.26 & 6.2 $\pm$ 1.2 & 6 \\
400-500 & 1.5 $\pm$ 0.6 & 0.48 $\pm$ 0.27 & 0.01 $\pm$ 0.01 & 0.28 $\pm$ 0.12 & 2.2 $\pm$ 0.7 & 2 \\
500-600 & 0.22 $^{+0.25}_{-0.22}$ & 0.11 $^{+0.12}_{-0.11}$ & 0.01 $\pm$ 0.01 & 0.18 $\pm$ 0.08 & 0.51 $\pm$ 0.29 & 0 \\
$>$600 & 0.06 $^{+0.07}_{-0.06}$ & 0.73 $\pm$ 0.44 & 0.02 $^{+0.03}_{-0.02}$ & 0.02 $^{+0.03}_{-0.02}$ & 0.84 $\pm$ 0.45 & 1 \\
\hline
\multicolumn{7}{c}{$\mtb \geq 175$\GeV, $\nj\geq5$, $N_\mathrm{t} \geq 1$, $\nb = 1$} \\
\hline
250-300 & 1.2 $\pm$ 0.5 & 0.30 $\pm$ 0.25 & 0.26 $\pm$ 0.21 & 0.02 $^{+0.03}_{-0.02}$ & 1.8 $\pm$ 0.6 & 0 \\
300-400 & 1.5 $\pm$ 0.7 & 0.34 $\pm$ 0.26 & 0.02 $\pm$ 0.01 & 0.14 $\pm$ 0.06 & 2.0 $\pm$ 0.8 & 0 \\
400-500 & 0.73 $\pm$ 0.40 & 0.20 $^{+0.22}_{-0.20}$ & 0.13 $^{+0.17}_{-0.13}$ & 0.04 $^{+0.05}_{-0.04}$ & 1.1 $\pm$ 0.5 & 1 \\
500-600 & 0.25 $\pm$ 0.22 & 0.54 $\pm$ 0.34 & 0.12 $^{+0.16}_{-0.12}$ & 0.10 $\pm$ 0.06 & 1.0 $\pm$ 0.4 & 4 \\
$>$600 & 0.15 $^{+0.33}_{-0.15}$ & 0.59 $\pm$ 0.49 & 0.07 $\pm$ 0.07 & 0.11 $\pm$ 0.05 & 0.92 $\pm$ 0.60 & 1 \\
\hline
\multicolumn{7}{c}{$\mtb \geq 175$\GeV, $\nj\geq5$, $N_\mathrm{t} \geq 1$, $\nb \geq 2$} \\
\hline
250-300 & 0.66 $\pm$ 0.26 & 0.11 $\pm$ 0.09 & 0.06 $\pm$ 0.05 & 0.09 $\pm$ 0.05 & 0.92 $\pm$ 0.29 & 3 \\
300-400 & 0.92 $\pm$ 0.39 & 0.12 $\pm$ 0.10 & 0.03 $\pm$ 0.03 & 0.14 $\pm$ 0.08 & 1.2 $\pm$ 0.4 & 3 \\
400-500 & 0.31 $\pm$ 0.17 & 0.03 $^{+0.04}_{-0.03}$ & $<$0.01 & 0.09 $\pm$ 0.06 & 0.43 $\pm$ 0.18 & 0 \\
500-600 & 0.30 $\pm$ 0.30 & 0.30 $\pm$ 0.21 & $<$0.01 & 0.09 $\pm$ 0.04 & 0.70 $\pm$ 0.37 & 0 \\
$>$600 & 0.13 $^{+0.29}_{-0.13}$ & 0.37 $\pm$ 0.32 & $<$0.01 & 0.12 $\pm$ 0.05 & 0.62 $\pm$ 0.43 & 1 \\
\hline
\end{tabular}
}
\end{table*}

\section{Search for top squarks in the single-lepton final state}
\label{sec:1lstop}

We also perform a search for top squarks in events with exactly one isolated electron or muon and considerable \MET.  The main SM backgrounds originating from \ttbar and \wjets~processes are suppressed using dedicated kinematic variables.  The dominant remaining backgrounds arise from lost-lepton processes and the surviving \wjets~background, both of which are estimated from control samples in data.

\subsection{Analysis strategy}
\label{sec1l:evtsel}

The search sample is selected using triggers that require either large \MET or the presence of an isolated electron or muon. The combined trigger efficiency for a selection of \MET$>250$\GeV and at least one lepton, as measured in a data sample with large $H_{\rm T}$, is
found to be $99\%$ with an asymmetric uncertainty of $^{+1}_{-3}\%$.
Selected events are required to have at least two jets with $\pt>30$\GeV, at least one of which must be \bq-tagged using the medium working point. We require exactly one well-identified and isolated electron or muon with $\pt > 20$\GeV, $\abs{\eta} < 1.442$ or  $< 2.4$, respectively, and $I_\text{rel} < 0.1$. Electrons in the forward region of the detector are not considered in this search due to a significant rate for a jet to be misidentified as an electron. To reduce the dilepton background originating from \ttbar and $\mathrm{tW}$ production, events are rejected if they contain a second electron or muon with $\pt > 5$\GeV and $I_\text{rel}<$ 0.2. A significant fraction of the remaining SM background originates from events with $\tauh$ decays. This contribution is reduced by rejecting events that have an isolated $\tauh$ candidate reconstructed using the HPS algorithm with $\pt > 20$\GeV and $\abs{\eta} < 2.4$. A further veto is placed on events containing isolated charged-hadron PF candidates with $\pt>10$\GeV and $\abs{\eta} < 2.5$. Candidates are categorized as being isolated if their isolation sum, \ie the scalar sum of the \pt of charged PF candidates within a fixed cone of $R = 0.3$ around the candidate, is less than 6\GeV and smaller than $10\%$ of the candidate \pt.

Single-lepton backgrounds originating from semileptonic \ttbar, \wjets, and single top quark processes are suppressed through the \MT of the lepton-neutrino system.
Background processes containing a single lepton from \W~boson decay have a kinematic endpoint for \MT at the \W~boson mass, modulo detector resolution and off-shell \W~boson mass effects.
In this analysis we require $\MT > 150$\GeV, which significantly reduces single-lepton backgrounds.
To further reduce the \ttbar background, we require the absolute value of the azimuthal angle between \ptvecmiss and the closest of the two highest-\pt jets, $\Delta\phi_{12}$, to be larger than 0.8, since the events that satisfy the \met and \MT requirements tend to have higher-\pt top quarks, and therefore smaller values of $\Delta\phi_{12}$ than signal events.

The remaining background after the preselection is dominated by dilepton events from \ttbar and $\topq\W$ production, where one of the leptons is not reconstructed or identified, and the
presence of the additional neutrino from the second leptonically decaying \W~boson makes it possible to satisfy the \MT requirement.

\begin{figure}[htb]
\centering
\includegraphics[width=0.48\textwidth]{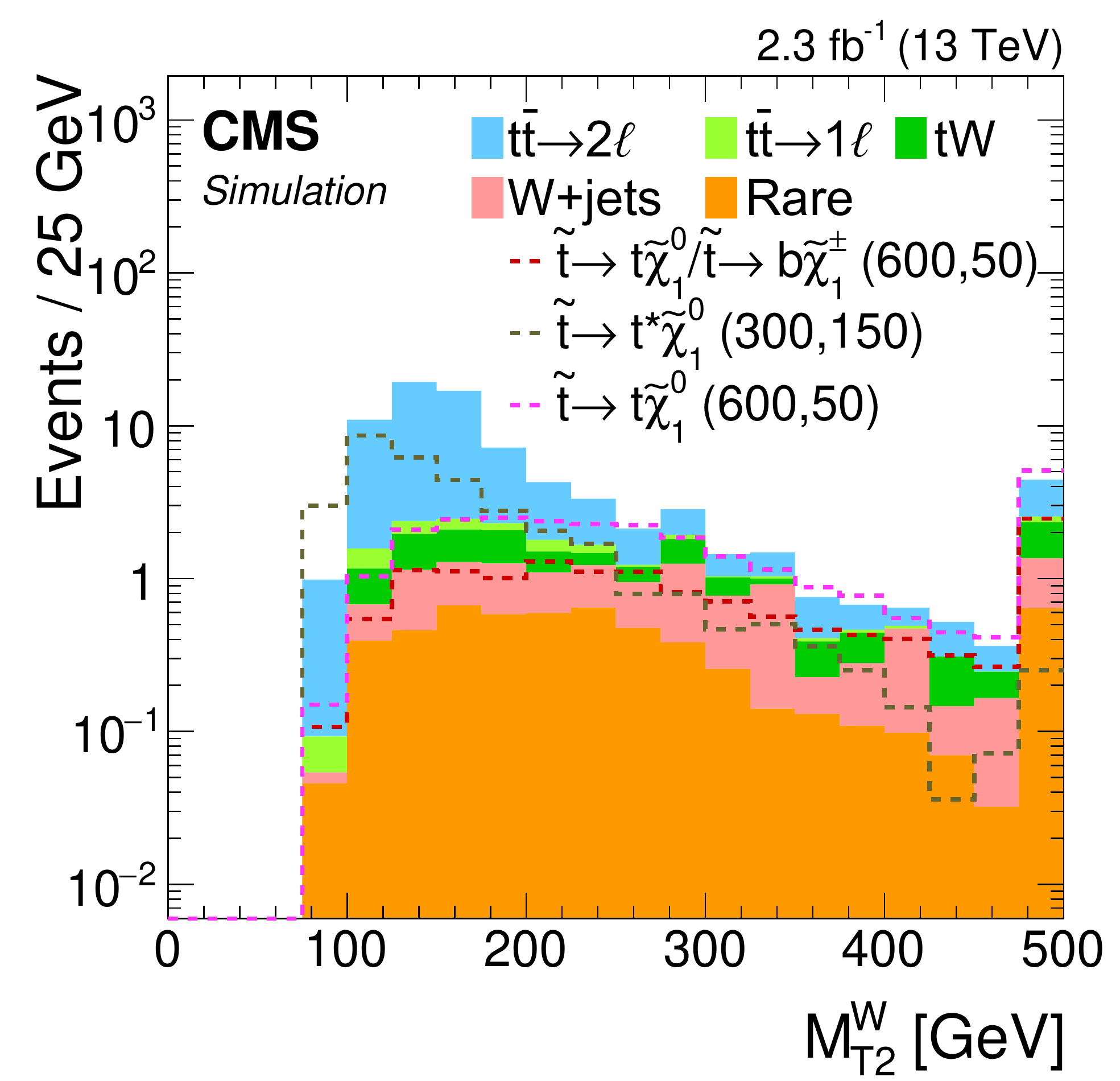}
\includegraphics[width=0.48\textwidth]{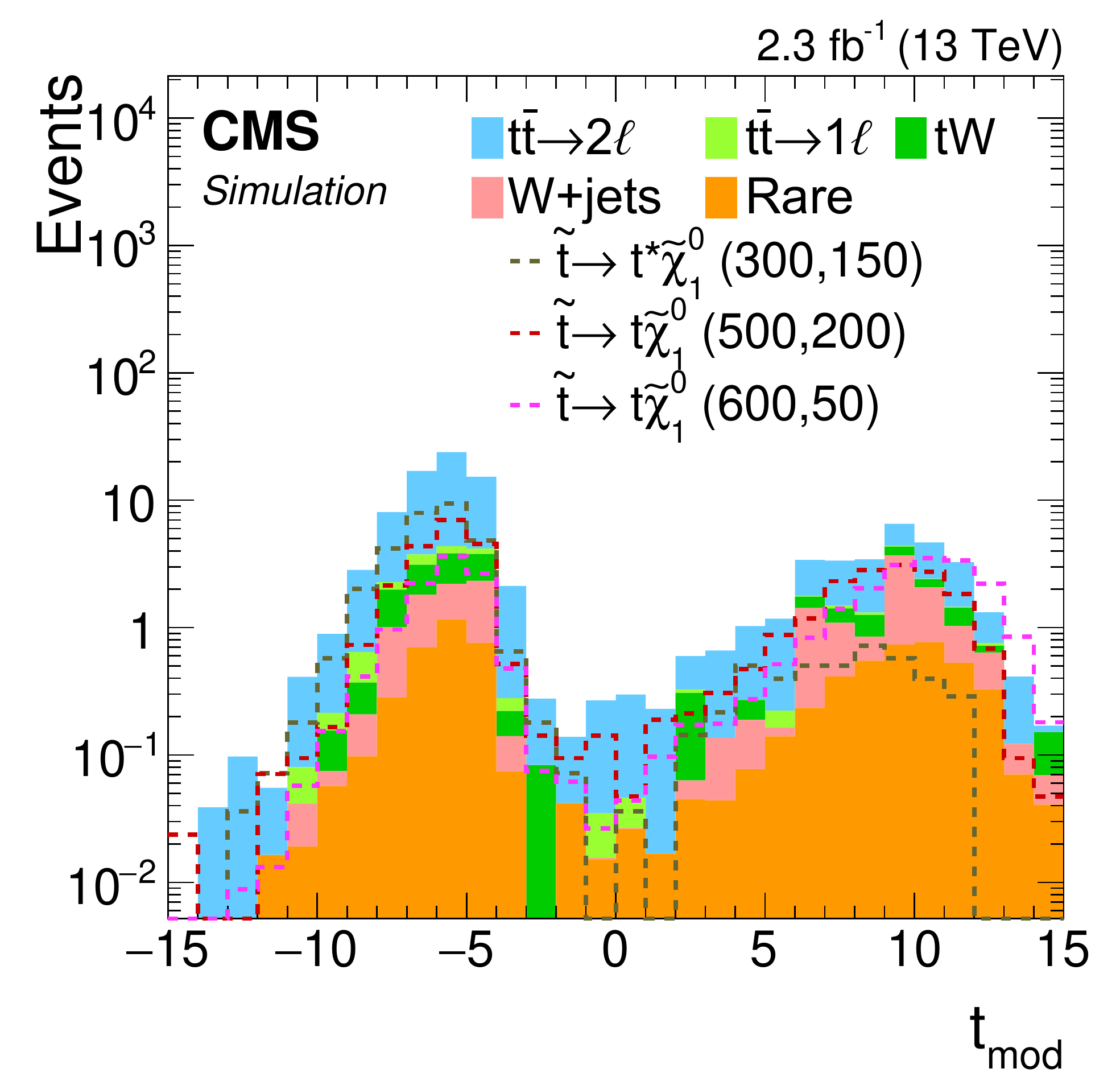}
\caption{\label{fig:mt2w_tmod}
{The \MTtW (\cmsLeft) and $t_\text{mod}$ (\cmsRight) distributions for signal and backgrounds after the preselection are shown. The \MTtW variable is shown for events with four or more jets,
while $t_\text{mod}$ is shown for events with at least two jets. Signal models with different top squark and neutralino mass hypotheses are shown for comparison.}}
\end{figure}

\label{sec1l:srdef}
Kinematic properties of signal events such as \met, \MT, and jet multiplicity depend on the decay modes of top squarks, as well as on the
mass splittings ($\Delta m$)  between the top squark, neutralino, and chargino (if present). As a basis for the search strategy in the topologies shown in Figs.~\ref{fig:diagram}(a) and~\ref{fig:diagram}(b), we require the presence of at least four jets.
Events are then categorized based on the value of the \MTtW variable~\cite{Bai:2012gs}, which is calculated for each event under the assumption that it originates from the dilepton \ttbar process with a lost lepton:
\begin{multline}\label{eq:mt2w}
\MTtW \equiv \mathrm{Min}\{m_\mathrm{y},\text{ consistent with: } \\
[p_\mathrm{1}^{2}=0,~(p_\mathrm{1}+p_{\ell})^2=p_\mathrm{2}^{2} = M_\mathrm{W}^{2},~\ptvec^{1} + \ptvec^{2}=\VEtmiss, \\
(p_\mathrm{1}+p_{\ell}+p_\mathrm{b_{1}})^{2}=(p_\mathrm{2}+p_\mathrm{b_{2}})^{2}=m_\mathrm{y}^{2} ]\},
\end{multline}
where $m_\mathrm{y}$ is the fitted parent particle mass, and $p_\mathrm{1},~p_{\ell},$ $p_\mathrm{2}$, $p_\mathrm{b_{1}},$ and $p_\mathrm{b_{2}}$ are the four momenta of the neutrino corresponding to the visible \W~boson decay, the lepton from the same decay, the \W~boson whose decay gives rise to the undetected lepton, and the two \bq~jet candidates, respectively.  To select the \bq~jet candidates, we examine all possible pairings with the three jets that have the highest CSV discriminator values.  The pairing that gives the lowest value of \MTtW defines the final estimate. The reconstruction of an event using the \MTtW variable helps discriminate signal from the dominant dilepton \ttbar background.
For large mass differences between the top squark and the neutralino, the $\MTtW>200$\GeV requirement significantly reduces the background while maintaining reasonable
signal efficiency. In contrast, for small-$\Delta m$ models, such a requirement results in a significant loss in signal efficiency.
To preserve sensitivity to both high- and low-$\Delta m$ scenarios, we subdivide the search sample into two event categories with $\MTtW>200$\GeV and $\leq200$\GeV.
The \MTtW distribution for events with at least four jets is shown in Fig.~\ref{fig:mt2w_tmod}~(\cmsLeft).

In signals with a large difference in mass between the top squark and the neutralino, a significant fraction of events can contain two quarks that merge into a single
jet as a result of the large boost of the top quark or \W~boson that decay into jets. These events would fail the four-jet requirement. To recover acceptance for such topologies,
 we define an additional SR in events with three jets. Since this region targets large $\Delta m$ signal scenarios, only events with
$\MTtW>200\GeV$ are considered.

To increase the sensitivity of this analysis to a mixed decay scenario (Fig.~\ref{fig:diagram}c) when the chargino and neutralino are nearly degenerate in mass,
SRs with exactly two jets are added. In events with low jet multiplicity the modified topness
variable ($t_\text{mod}$)~\cite{Graesser:2012qy} provides improved dilepton \ttbar\ rejection:
\ifthenelse{\boolean{cms@external}}{
\begin{equation} \label{eq:modtopness}
\begin{aligned}
t_\text{mod} = \ln(\min S),\quad\text{with}\\
S(\vec{p}_{\PW}, p_{z},\nu) =& \frac{(m_{\PW}^2-(p_\nu+p_{\ell})^2)^2}{a_{\PW}^4} \\
&+
\frac{(m_{\PQt}^2 - (p_{\PQb}+p_{\PW})^2)^2}{a_{\PQt}^4}.
\end{aligned}
\end{equation}
}{
\begin{equation} \label{eq:modtopness}
t_\text{mod} = \ln(\min S),
\text{ with }
S(\vec{p}_{\PW}, p_{z},\nu) = \frac{(m_{\PW}^2-(p_\nu+p_{\ell})^2)^2}{a_{\PW}^4} +
\frac{(m_{\PQt}^2 - (p_{\PQb}+p_{\PW})^2)^2}{a_{\PQt}^4}.
\end{equation}
}
This equation uses the mass constraints for the particles and also the assumption that $\ptvecmiss=\vec{p}_\mathrm{T,W}+\vec{p}_{\mathrm{T},\nu}$.  The first term constrains the \W~boson whose lepton decay product is the detected lepton, while the second term constrains the top quark for which the lepton from the \W~boson decay is lost in the reconstruction.
Once again, we consider all possible pairings of \bq~jet candidates with up to three jets with highest CSV discriminator values.
The calculation of modified topness uses the resolution parameters $a_\mathrm{W} = 5$\GeV and $a_\mathrm{t} = 15$\GeV, which determine the relative weighting of the mass shell conditions.
We select events with $t_\text{mod}>6.4$. The definition of topness used in this analysis is modified from the one originally proposed
in Ref.~\cite{Graesser:2012qy}: namely, the terms corresponding to the detected leptonic top quark decay and the centre-of-mass energy are dropped since in events with low jet multiplicity the second \bq~jet is often not identified.
In these cases, the discriminating power of the topness variable is reduced when a light-flavour jet is used instead in the calculation.
The modified topness is more robust against such effects and provides better signal sensitivity in these SRs than the $\MTtW$ variable.
The distribution of modified topness for events with at least two jets is shown in Fig.~\ref{fig:mt2w_tmod}~(\cmsRight).

Finally, events in each of the categories described above are further classified into different SRs
based on the value of \met. This results in a total of nine exclusive SRs as summarized in Table~\ref{tab1l:SR}.
\begin{table*}
\centering
\topcaption{\label{tab1l:SR} Summary of the SR definitions for the single-lepton search.}
\begin{tabular}{lrrr|rrr}
\hline
Targeted models & $\nj$ & $M_\mathrm{T2}^{\PW}$ [\GeVns{}] & $t_\text{mod}$ & \multicolumn{3}{c}{$E_\mathrm{T}^\text{miss}$ [\GeVns{}]} \\
\hline
Low-${\Delta}m$ & $\geq4$ & $\leq 200$ & & $250$--$325$ & $>325$ & \\
High-${\Delta}m$ &  $\geq4$ & $> 200$ & & $250$--$350$ & $350$--$450$ & $>450$ \\
\hline
Boosted high-${\Delta}m$ & $=$3 & $>$200 & & 250--350 & $>$350 & \\
\hline
Degenerate $\PSGcpmDo$ and $\PSGczDo$ & $=$2 & & $>$6.4 & 250--350 & $>$350 & \\
\hline
\end{tabular}
\end{table*}

\subsection{Background estimation}
\label{Sec1l:BkgEst}

Three categories of backgrounds originating from SM processes remain after the preselection described in Section~\ref{sec1l:evtsel}.
The dominant contribution arises from backgrounds with a lost lepton, primarily from the dilepton \ttbar process.
 A second class of background events originates from SM processes with a single leptonically decaying \W~boson. Preselection
 requirements of $\met>250\GeV$ and $\MT>150\GeV$ strongly suppress this background. The suppression is much stronger
for events with a \W~boson originating from the decay of a top quark than for direct \W~boson production, as the mass of the
top quark imposes a constraint on $M_{\W}$. As a result, large values of \MT in semileptonic \ttbar events are
dominated by \met resolution effects, while for events in which the \W~boson is produced directly (\wjets) they are mainly a function of the width of the \W~boson.
The third class of background events includes rare SM processes such as \W\Z~and $\ttbarZ$ (where the
\Z~boson decays to neutrinos), with smaller contributions from $\ttbarW$, $\ttbar\gamma$, and processes with two or three
electroweak vector bosons. The QCD background is negligible in this
search due to requirements on the presence of a high-\pt isolated lepton, large \met, and large \MT.

\subsubsection{Lost-lepton background}\label{sec1l:dilepton}

The lost-lepton background is estimated from data in dilepton CRs, where we require the presence of a second lepton passing the rejection requirements but with $\pt>10\GeV$, an isolated track, or a $\tauh$ candidate. This is done again by extrapolating the data in the dilepton CRs to the SRs using  transfer factors obtained from simulation.  We use the same preselection requirements on $\met$ and $\MT$ as in the search regions.
We remove the subdivision in \met and the separation of the three and at least four jet regions to increase the statistical power of the CRs, and arrive at three CRs: exactly two jets and $t_\text{mod}>6.4$, at least three jets and $\MTtW\leq200$\GeV, and at least three jets and $\MTtW>200$\GeV. These control regions have a purity in dilepton events of $>$ 97\%. Additional transfer factors are therefore needed to account for the extrapolation in jet multiplicity and \met requirements; these are derived from simulation. The background estimate can be written as follows:
\ifthenelse{\boolean{cms@external}}{
\begin{equation}\label{eq:diLep_SR_est_full}
\begin{aligned}
N^\text{pred}_\mathrm{LL} &= N^\text{data}_{2\ell}~T_\text{LL}~T_{\met,\nj}, \\
T_\text{LL} &= \frac{N^\text{sim}_{1\ell}}{N^\text{sim}_{2\ell}}, \quad T_{\met,\nj} =  \frac{N^\text{sim}_{1\ell}(\met,\nj)}{N^\text{sim}_{1\ell}},
\end{aligned}
\end{equation}
}{
\begin{equation}\label{eq:diLep_SR_est_full}
N^\text{pred}_\mathrm{LL} = N^\text{data}_{2\ell}~T_\text{LL}~T_{\met,\nj}, \quad
T_\text{LL} = \frac{N^\text{sim}_{1\ell}}{N^\text{sim}_{2\ell}}, \quad T_{\met,\nj} =  \frac{N^\text{sim}_{1\ell}(\met,\nj)}{N^\text{sim}_{1\ell}},
\end{equation}
}
where  $N^\text{data}_{2\ell}$ is the number of events observed in data in the dilepton CR.  The largest systematic uncertainty in the background estimate is due to the statistical uncertainties of the event yields in data CRs and the estimates from simulated samples (10--30\%).  The signal contamination in this CR is around 10\% for the bulk of the studied parameter space and is taken into account in the final interpretation.
The transfer factor $T_\text{LL}$ is obtained from simulation, and estimates the probability that a lepton is not identified in the detector, accounting for the kinematic acceptance and the efficiency of the lepton selection criteria.
The second transfer factor, $T_{\met,\nj}$, extrapolates the inclusive estimate to individual SR bins.  This transfer factor, also obtained from simulation, is validated by checking the modelling of the jet multiplicity and of the \met spectrum in dedicated data CRs, which will be described in the following paragraphs.

The dilepton \ttbar background contributes to the SRs with three or more jets only if jets from ISR or final-state radiation (FSR) are also present, or when a $\tau_h$ decay is misidentified as an additional jet. The modelling of jet multiplicity is checked in a high-purity dedicated dilepton data control sample with one electron and one muon, at least two \bq-tagged jets, and $\met > 250$\GeV.
The differences between data and simulation are used to estimate scale factors relative to the baseline selection of events with at least two jets.  The scale factors are $1.10\pm0.06$ for three-jet events and $0.94\pm0.06$ for events with at least four jets.  Within statistical uncertainties, these factors display no \met dependence.  The scale factors are applied to the dilepton \ttbar simulation when extrapolating the inclusive background prediction into the specified jet multiplicity bins. The statistical uncertainties in these scale factors are also propagated to the predictions in the SRs. The uncertainty in the modelling of the jet multiplicity ranges up to 3\%.

The extrapolation in \met is carried out through simulation, and it must be verified that its resolution is accurately modelled.
Changing the resolution can lead to a different \met spectrum. In this analysis we are interested in the effect of the \met resolution in events containing intrinsic \met because of the presence of neutrinos in the events.
This effect is estimated by comparing a \gjets sample in data with simulation.
The events are selected using a single-photon trigger with $\pt> 165$\GeV and $\abs{\eta} < 2.4$. Photons are required to pass stringent identification criteria.
We use the photons to mimic the neutrinos in the event, with the photon momentum serving as an estimate of the sum of the neutrino momenta.

The photon \pt spectrum in data and in simulation is reweighted to match that of the neutrinos in the background-simulation sample.
For dilepton \ttbar events, this corresponds to the $\nu\nu$-$\pt$ spectrum.  To model the \MET resolution, the transverse momentum of the photon system is added vectorially to the \ptvecmiss and the resulting \met spectrum is compared between data and simulation.  We use this modified \MET definition to calculate our discriminants. For this CR, we then apply selection criteria close to the SR criteria, except that selections related to the lepton are dropped, the presence of a well-identified photon is required, and the requirement of a \bq-tagged jet is reversed so as to suppress effects related to semileptonic heavy-flavourdecays. Corrections for the observed differences, which can go up to 15\%, are applied to events in the simulated samples and the uncertainties propagated to the final background estimate, resulting in an uncertainty of 1--4\% in the lost-lepton background prediction.

\subsubsection{One-lepton background}\label{sec1l:onelepton}
In SRs with a high \MTtW or modified topness requirement, the \wjets background is estimated using a data control sample containing no \bq-tagged jets. For SRs with a low-\MTtW requirement, this background constitutes less than 10\% of the total SM background. In these SRs we do not employ an estimate based on data, but instead use the \wjets~background estimate directly from simulation. The semileptonic \ttbar background is also estimated from simulation.

The CRs used to extract the \wjets background in the SRs with a high \MTtW or modified topness requirement are again not subdivided in \met to have a sufficient number of events to carry out the prediction. We therefore use three CRs for this background estimate: exactly two jets with $t_{\text{mod}} >$ 6.4, exactly three jets with \MTtW $>$ 200\GeV, and at least four jets with $\MTtW > 200$\GeV.
We extrapolate the yields from the CRs to the SRs by applying transfer factors from simulation for the extrapolation in \met and number of \bq-tagged jets:
\begin{equation}
N^\text{pred}_{\wjets} = (N^\text{data}_{\nb = 0} - N^\text{non-\wjets}_{\nb = 0})~T_{\met}~T_{\nb},
\end{equation}
with $N^\text{data}_{\nb = 0} - N^{\text{non-\wjets}}_{\nb = 0}$ representing the event yield in the CR after subtracting the estimated contribution from other SM background processes. The non-1$\ell$ contribution in the CRs, $N^{\text{non-\wjets}}_{\nb = 0}$, is estimated from simulation and amounts to roughly 25--35\%.  A 50\% uncertainty is assigned to the subtraction. The largest source of uncertainty is again the limited size of the data and simulation samples.  The statistical uncertainty of these samples results in an uncertainty of 20--40\% in the \wjets background estimate.

 The transfer factor $T_{\met}$ extrapolates the yields from the inclusive CR with $\met>50$\GeV to the exclusive \met regions.   The main uncertainties in this extrapolation factor can be attributed to the modelling of the neutrino \pt spectrum, the \W~boson width, and the \met resolution.    The neutrino \pt spectrum is checked in a data sample enriched in \wjets, with no \bq-tagged jets and $60<\MT<120$\GeV.  No large mismodelling of \met is observed.  Therefore, we do not apply any corrections to the neutrino \pt spectrum but only propagate the statistical limitation of this study as the uncertainty (6--22\%) in the modelling of the neutrino \pt spectrum. The uncertainty in the \W~boson width (3\%~\cite{PDG}) is estimated by scaling the four-vectors of the \W~boson decay products appropriately. The \met resolution effects on this background are studied using the same method as described in Section~\ref{sec1l:dilepton}, giving rise to a 1--3\% uncertainty.

The other transfer factor, $T_{\nb}$, performs the extrapolation in the number of \bq-tagged jets for each \met bin.  Scale factors are applied to the simulation to match the \bq~tagging efficiency in data.  The largest uncertainty in this transfer factor is the fraction of the heavy-flavour component in the \wjets sample; we assign a 50\% uncertainty to this component.  We performed a dedicated cross-check in a CR with one or two jets and at least 50\GeV of \met.  Data and simulation were found to be in agreement in the \bq~jet multiplicity within uncertainties.
After taking into consideration the additional sources of systematic uncertainty described in Section~\ref{sec:systematics}, the total uncertainty in the \wjets  estimate varies from 50\% to 70\%.

The semileptonic \ttbar background is never larger than 10\% of the total background estimate. We rely on simulation to estimate it. The main source of uncertainty in this estimate is the modelling of the \met resolution because poor resolution can enhance the contributions at large \MT. The studies of \met resolution presented in Section~\ref{sec1l:dilepton} indicate that it could be mismodelled by about 10\% in simulation.  Changes in the simulated \met resolution by a corresponding amount provide an uncertainty of 100\% in the semileptonic \ttbar estimate.

\subsubsection{Rare standard model backgrounds}
The ``rare'' background category includes $\ttbar$ production in association with a vector boson (\W, \Z, or $\gamma$), diboson, and triboson events.
Within this category, \W\Z~events dominate the SRs with two jets, and $\ttbarZ$ events with the \Z~boson decaying into a pair of neutrinos ($\znunu$) dominate regions of higher jet multiplicity.
The expected contributions from these backgrounds are small, and the simulation is expected to model the kinematics of these processes well in the regions of phase space relevant to the SRs. The rare backgrounds are therefore estimated using simulation. We assess the theoretical and experimental uncertainties affecting the estimates as described in Section~\ref{sec:systematics}, resulting in a total uncertainty of 15--26\%, depending on the SR.

\subsection{Results}

\begin{figure}[htb]
\centering
\includegraphics[width=\cmsFigWidth]{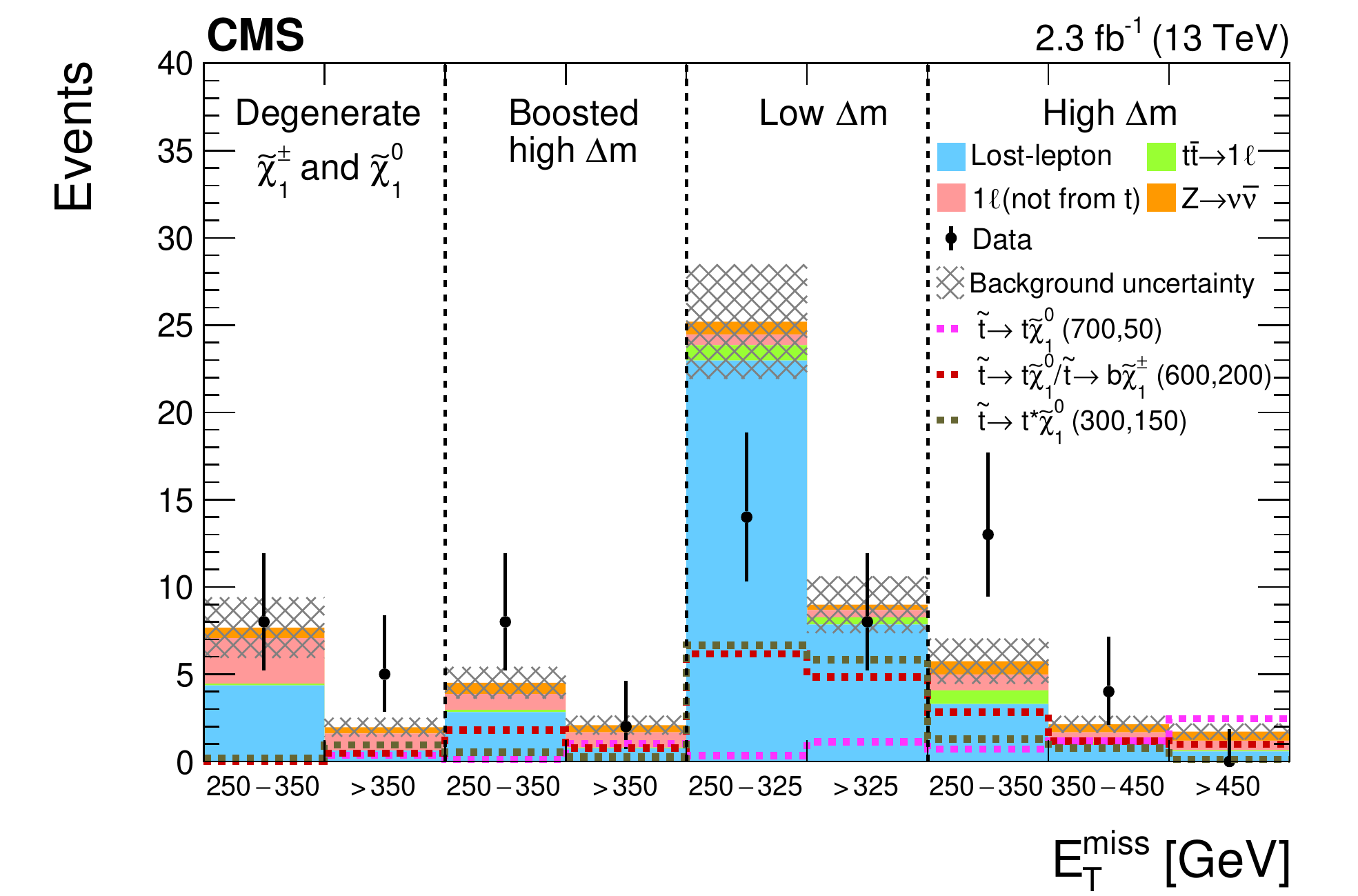}
\caption{\label{fig1l:results}  Background estimates from data and simulation, together with the observed yields in the SRs of the single-lepton analysis, described in Table~\ref{tab1l:SR}. The uncertainties, which are the quadratic sums of statistical and systematic uncertainties, are indicated by the cross-hatched areas. The SM background predictions shown do not include the effects of the maximum likelihood fit to the data.  Three signal hypotheses are overlaid. The hypothesis $\stopq\to\topq\lsp/\stopq\to\bq\PSGcpmDo$ has branching fractions $\mathcal{B}(\stopq\to\topq\lsp)=\mathcal{B}(\stopq\to\bq\PSGcpmDo)=0.5$.}
\end{figure}

The background expectations and the corresponding yields for each SR are summarized in Table~\ref{tab1l:results} and  in Fig.~\ref{fig1l:results}. Overall, the observed and predicted yields agree within two standard deviations in all SRs.
For signals of top squark pair production for different mass hypotheses, the maximum observed significance obtained by combining the results in different SRs is 1.2 standard deviations for a top squark mass of $\approx$400\GeV and a massless LSP hypothesis. We therefore find no evidence for top squark pair production.

\begin{table*}[htb]
\centering
\topcaption{\label{tab1l:results} Background estimates from data and simulation, and observed data yields for the single-lepton top squark analysis using 2.3\fbinv  of data collected during 2015 pp collisions. The uncertainties are the quadratic sums of statistical and systematic uncertainties.}
\begin{tabular}{rr@{\,$\pm$\,}lr@{\,$\pm$\,}lr@{\,$\pm$\,}lr@{\,$\pm$\,}lr@{\,$\pm$\,}lr}
\hline
\multirow{2}{*}{\met [\GeVns{}]} &  \multicolumn{2}{c}{\multirow{2}{*}{Lost-lepton}} & \multicolumn{2}{c}{$1\ell$ (not}  &  \multicolumn{2}{c}{\multirow{2}{*}{$\ttbar\to\
1\ell$}} &  \multicolumn{2}{c}{\multirow{2}{*}{$\cPZ\to\cPgn\cPagn$}} & \multicolumn{2}{c}{Total} & \multirow{2}{*}{Data} \\
 & \multicolumn{2}{c}{~} & \multicolumn{2}{c}{from top)} & \multicolumn{2}{c}{~} & \multicolumn{2}{c}{~} & \multicolumn{2}{c}{background} &  \\
\hline
 & \multicolumn{11}{l}{Degenerate $\PSGcpmDo$ and $\PSGczDo$: $2$ jets, $t_\text{mod}>6.4$} \\
\hline
$250$--$350$    & 4.4&1.4 & 2.61&0.99 & 0.09&0.09 & 0.60&0.12 & 7.7&1.7 & 8 \\
$>$350    & 0.62&0.23 & 0.98&0.47 & \multicolumn{2}{c}{$<$0.03} & 0.36&0.13 & 1.96&0.54 & 5 \\
\hline
 & \multicolumn{11}{l}{Boosted high $\Delta m$: $3$ jets, $\MTtW>200\GeV$} \\
\hline
$250$--$350$    & 2.83&0.73 & 0.92&0.52 & 0.12&0.12 & 0.64&0.13 & 4.51&0.91 & 8 \\
$>$350    & 0.74&0.21 & 0.88&0.50 & 0.05&0.05 & 0.41&0.09 & 2.08&0.55 & 2 \\
\hline
 & \multicolumn{11}{l}{Low $\Delta m$: $\geq4$ jets, $\MTtW\leq200\GeV$} \\
\hline
$250$--$325$ & 23.0&3.2 & 0.61&0.61 & 0.88&0.88 & 0.74&0.17 & 25.2&3.4 & 14 \\
$>$325 & 7.9&1.5 & 0.45&0.45 & 0.40&0.40 & 0.30&0.11 & 9.0&1.6 & 8 \\
\hline
 & \multicolumn{11}{l}{High $\Delta m$: $\geq4$ jets, $\MTtW>200\GeV$} \\
\hline
$250$--$350$ & 3.29&0.91 & 0.92&0.46 & 0.78&0.78 & 0.76&0.19 & 5.8&1.3 & 13 \\
$350$--$450$ & 0.94&0.27 & 0.54&0.34 & 0.18&0.18 & 0.46&0.14 & 2.13&0.48 & 4 \\
$>$450 & 0.57&0.21 & 0.55&0.36 & 0.07&0.07 & 0.52&0.17 & 1.71&0.45 & 0 \\
\hline
\end{tabular}
\end{table*}

\section{Search for pair production of bottom squarks or of top squarks decaying to charm quarks}
\label{sec:sbottom}

This search is motivated by the production
of pairs of bottom or top squarks, in which each $\sbottomq$ or $\stopq$  decays, respectively, into a bottom or a charm quark
and a neutralino.
In the latter search, the difference between the $\stopq$ and  $\lsp$ masses is assumed to be less than 80\GeV, and the only top squark decay mode considered is through a flavour changing neutral current to $\cq\lsp$. Small
mass splittings $\Delta m = m_{\stopq}-m_{\lsp}$ or $\Delta m = m_{\sbottomq}-m_{\lsp}$  between the top or bottom squark and the
neutralino leave little visible energy in the detector, making signal events difficult to distinguish from
SM background. However, events with an energetic ISR jet recoiling against the \ptvecmiss originating from the neutralino can provide a distinct topology for signals with compressed mass spectra, \ie with small $\Delta m$. We thus perform a search for
events with an ISR jet and significant \met.

\subsection{Analysis strategy}
\label{sec:sbottomsel}

Events in the search sample are recorded using the same trigger as that for the top squark search in the all-jet final state, requiring the presence of large \met and at least two energetic jets within the tracker acceptance.  After applying an offline selection requiring $\met>250$\GeV and at least two
jets with $\pt>60$\GeV, we find the trigger efficiency to be greater than 97\%.
We veto events that have at least four jets with \pt above 50\GeV.
The veto and its threshold are motivated by the harder \pt spectrum of the fourth jet in semileptonic \ttbar events compared to the signal, in which extra jets originate from ISR or FSR.
To reduce the SM background from processes with a leptonically decaying \W~boson, we reject events containing isolated electrons or muons with $I_\text{rel}<0.1$ and $\abs{\eta} < 2.5$, or $I_\text{rel}<0.2$ and $\abs{\eta} < 2.4$, respectively, and with $\pt > 10$\GeV. The contribution containing $\tauh$ decays is reduced by placing a veto on events containing charged-hadron PF candidates with $\pt>10$\GeV, $\abs{\eta} < 2.5$, and an isolation sum smaller than $10\%$ of the candidate \pt.

The dominant SM background sources are \zjets production with \znunu, and the lost-lepton background originating from \wjets, \ttbar, and single top quark processes with leptonic \W~boson decays.
A smaller background contribution comes from QCD events in which large \met originates from jet mismeasurements and the direction of \ptvecmiss is often aligned with one of the jets. To suppress this background we require that the absolute difference in azimuthal angle between the \ptvecmiss and the closest of the three leading jets ($\dphijonetwothree$) is greater than 0.4. Two sets of SRs are defined to optimize the sensitivity for signal models with either compressed or noncompressed mass spectra.

In addition to the criteria discussed above, for regions targeting noncompressed scenarios we require that the \pt of the leading jet be above 100\GeV
and that the event contain at least one additional jet with \pt above 75\GeV. We also require that the two highest-\pt jets be identified as \bq~jets. These requirements suppress events originating from \W~and \Z~boson production, for which the leading jets have a softer \pt spectrum since they are produced by ISR or FSR.
To maintain a stable \bq~tagging efficiency as a function of jet \pt, both the loose and medium working points of the \bq~tagging algorithm are used to identify \bq~jets.  The \bq~tagging efficiency of the medium working point depends strongly on the jet \pt and degrades by about 20--30\% for jets with \pt above 500\GeV, while the efficiency of the loose working point is more stable with increasing jet \pt. Specifically, we use the loose working point to identify \bq-tagged jets when the leading jet has \pt above 500\GeV, and the medium working point otherwise.  Since such high-\pt jets are less likely to occur in SM processes, the higher misidentification rate of the loose working point results in only a small increase in the SM background.

{\sloppypar
The distribution of $\minMT \equiv \min[\MT(\text{j}_{1}, \ptvecmiss), \MT(\text{j}_{2}, \ptvecmiss)]$, where $\text{j}_{1}, \text{j}_{2}$ are the two highest-\pt jets, is expected to have a kinematic endpoint
at the mass of the top quark when \ptvecmiss and the closest jet originate from the semileptonic decay of a top quark. In the noncompressed search sample we require \minMT to be greater than 250\GeV.
Events in this sample are then categorized by \HTonetwo, defined for the purposes of this analysis as the scalar sum of the \pt of the two leading jets, and the \mct kinematic variable. The boost-corrected cotransverse mass~\cite{MCT,MCT1}, \mct, is defined by:
\par}
\begin{equation}
\mct^2(\mathrm{j_1, j_2})=  2\pt(\mathrm{j_1})\pt(\mathrm{j_2}) [1+\cos\Delta\phi(\mathrm{j_1,j_2})].
\end{equation}

For scenarios in which two particles are pair-produced and have the same decay chain, the \mct distribution has an endpoint determined by the masses of the parent and decay-product particles.  For $\sbottomq \to \bq\lsp$ this endpoint is at $(m(\sbottomq)^{2}-m(\PSGczDo)^2)/m(\sbottomq)$.

For signals with compressed mass spectra,
high-\pt ISR is required to be able to reconstruct the quarks as jets and obtain a large value of \met.  Compressed SRs require therefore a leading jet with $\pt>250$\GeV that is back-to-back relative to the \ptvecmiss ($\dphijmet>2.3$).   Since such ISR jets are not expected to originate from \bq~quarks, we require that the leading jet fail the loose \bq-tagging requirement.

We relax the thresholds on the second jet \pt and on the \minMT to 60 and 200\GeV, respectively, and categorize events in the search sample according to the number of \bq-tagged jets. The \mct observable loses its discriminating power for these compressed signal models due to the small mass splitting
between the parent particle and $\lsp$.
The \met is therefore used as the main discriminant, with different \met thresholds applied to define the final SRs.

The baseline selections for both noncompressed and compressed regions are summarized in Table~\ref{tab:selection}, while the definitions of the two sets of SRs are described in Table~\ref{tab:noncompandcomp}.

\begin{table*}[!ht]
\centering
\topcaption{A summary of the baseline selections used for the noncompressed and compressed $\sbottomq$ and $\stopq \to \cq \lsp$ compressed SRs.}
\begin{tabular}{lll}
\hline
   Selection&   Noncompressed & Compressed    \\ \hline
$\nj$ & 2$\leq\nj\leq$3 & 2$\leq\nj\leq$3 \\
First jet \pt &$> 100$\GeV  &$> 250$\GeV \\
Second jet \pt &$> 75$\GeV  &$> 60$\GeV \\
Veto fourth jet & \pt $>$ 50\GeV & \pt $>$ 50\GeV \\
Lepton and isolated track veto &\pt$>$ 10\GeV& \pt$>$ 10\GeV \\
\bq~tagging & First and second jets are \bq-tagged & Leading jet is not \bq-tagged \\
\met & $>$250\GeV& $>$250\GeV \\
$\dphijonetwothree$ & $>$0.4 &  $>$0.4 \\
\dphijmet& \NA &  $>$2.3\\
\minMT&$>$ 250\GeV& $>$200\GeV \\
\HTonetwo&$>$ 200\GeV& \NA \\
\mct&$>$250\GeV& \NA \\
\hline
\end{tabular}
\label{tab:selection}
\end{table*}

\begin{table*}[!htb]
\centering
\topcaption{\label{tab:noncompandcomp}  The categorization in \HTonetwo\ and \mct for the SRs targeting noncompressed signal scenarios, and in $\nb$ and \met for those targeting compressed signal scenarios.}
{\begin{tabular}{cc c c c c c c}
\multicolumn{1}{c}{} & \multicolumn{7}{c}{Noncompressed SRs}
\\
\hline
\HTonetwo [\GeVns{}] & \multicolumn{7}{c}{\mct [\GeVns{}]} \\
\hline
250--500 & & 250--350 & & 350--500 & & $>$500 & \\
$>$500  & & 250--350 & & 350--500 & & $>$500 & \\
\hline
\\[-0.5ex]
\multicolumn{1}{c}{} & \multicolumn{7}{c}{Compressed SRs}
\\
\hline
$\nb$ & \multicolumn{7}{c}{\met [\GeVns{}]} \\
\hline
0 & 250--350 & 350--450 & 450--550 & 550--700 & 700--850 & 850--1000 & $>$1000 \\
1 &
250--350 & 350--450 & 450--550 & 550--700 & $>$700  &   &  \\
2 & $>$250  &   &  &   &   &  &  \\
\hline
\end{tabular}
}
\end{table*}

\subsection{Background estimation}
\label{sec:sbottombackgrounds}

The SM background contributions originating from \znunu, lost-lepton, and QCD processes are estimated from dedicated data CRs as discussed below. Smaller contributions from other SM processes, such as diboson (VV) processes, are estimated from simulation, and an uncertainty of 50\% is assigned to these contributions.

\subsubsection{Estimation of the \texorpdfstring{\znunu}{Z nu nu}~background}
\label{sec:zinv_dy}

The \znunu\ background is estimated from a high-purity data sample of \zmm~events in which we remove the muons and recalculate the relevant kinematic variables to emulate \znunu\ events. The trigger used to collect this CR requires the presence of a high-\pt muon with $\abs{\eta} < 2.1$. In keeping with the trigger constraints, the sample is selected by requiring the presence of two isolated muons in the event with $\pt> 50\,(10)$\GeV and
$\abs{\eta}<2.1$\,(2.4) for the leading (trailing) muon. The invariant mass of the dimuon pair is required to be within 15\GeV of the \Z~boson mass~\cite{PDG}. Each muon is required to be separated from jets in the event by $\Delta R > 0.3$.

Apart from the lepton selection, we apply the same object and event selection criteria as described in
Section~\ref{sec:sbottomsel} to this sample, with the exception that \bq~jets are selected using the loose working point of the \bq~tagging algorithm to improve the statistical power of the data CR. Events in the selected sample are subdivided into CRs corresponding to the noncompressed and compressed SRs.
 The observed events in these data CRs, $N^\text{data}_{\mu\mu}$, are translated into an estimation of the \znunu\ contribution in the SRs with the help of simulation, as follows:
\begin{equation}
N^\text{pred}_{\znunu} = \frac{N^\text{data}_{\mu\mu}-N^{\text{non-\Z}}_{\mu\mu}}{A~\epsilon}~~R_{\rm Z}^{\mu\mu\to\nu\nu}~~\kappa,
\label{equ:Zvv}
\end{equation}
where $N^{\text{non-}\Z}_{\mu\mu}$, representing the small contamination in the CRs due to  \ttbar, \wjets, single top quark, and diboson processes, is estimated from simulation.
The corrected dimuon event yield is scaled by the kinematic and detector acceptance of muons from \Z~bosons, $A$, and the muon reconstruction, identification, and isolation efficiency $\epsilon$.  The acceptance and efficiency are determined from simulation.  Efficiency scale factors are applied to correct for differences between data and simulation.  These scale factors are determined with a ``tag-and-probe" method in \zmm\ events~\cite{Chatrchyan:2012xi}.
The product of the muon acceptance and efficiency, $A\epsilon$, varies from 0.6 in the low-\mct and low-\met regions to 0.9 in the high-\mct and high-\met regions.
The correction factor $R_{\rm Z}^{\mu\mu\to\nu\nu} = 5.942\pm0.019$~\cite{PDG} represents the ratio of the \Z~boson branching fractions to neutrinos and leptons.
The remaining term, $\kappa$, accounts for differences in the \bq~tagging efficiency and misidentification rate between the CRs and SRs, resulting from the use of different \bq~tagging working points.
These $\kappa$ factors are determined from \zll~simulation and corrected for known differences in the performance of the \bq~tagging algorithm between data and simulation
as measured in samples of multijet and \ttbar events~\cite{CMS-PAS-BTV-15-001}.
The value of the \bq~tagging $\kappa$ factor ranges from 0.10 to 0.15 for the noncompressed SRs, and from 0.20 to 0.25 for the $\nb = 1$ compressed SRs, while it is about 0.15 for the $\nb = 2$ compressed SR.

The largest uncertainty in the \znunu\ background estimate arises from the limited event yields in the dimuon CR, corresponding to a 10--100\% uncertainty in the \znunu\ prediction.
We correct for the estimated contributions to the CR from SM processes other than \zmm\ using simulation samples with an assigned uncertainty of 50\% in their normalization.  This leads to an uncertainty of 2--20\% in the background estimate.
Other experimental and theoretical sources of uncertainty, to be discussed in Section~\ref{sec:systematics}, result in an additional 2--8\% uncertainty in $A\epsilon$, and a 2\% uncertainty is assigned in
all SRs to account for the uncertainty in the \Z~boson branching fractions.
The uncertainty in the \bq~tagging $\kappa$ factors is assessed by varying the data-to-simulation \bq~tagging correction factors according to their measured uncertainties.
Additionally, the dependence of $\kappa$ on the heavy-flavour content in \Z~boson events is evaluated by varying the $\cPZ+$\bbbar and $\cPZ+$\ccbar
fractions in simulation by 20\% based on the uncertainty in the CMS $\cPZ+$\bbbar measurement~\cite{Zbb}, resulting in an additional uncertainty of 10--20\% in the \znunu\ estimate.

\subsubsection{Estimation of the lost-lepton background}
\label{sec:lostlep}

The lost-lepton background in each SR is estimated from a single-lepton CR in data selected by inverting the electron and muon vetoes in events collected with the same trigger as used to record the signal sample.
We relax the \bq~tagging requirement in the CRs using the loose working point in the noncompressed selection,
while keeping the same requirement as in the SRs for the compressed regions.  In all other respects, the CRs are defined through the same selection criteria as the corresponding SRs, including requirements on the \HTonetwo, \mct, $\nb$, and \MET, to remove any dependence of the prediction on the modelling of these kinematic variables in simulation. The possible contamination from signal in the single-lepton CR is negligible, less than 1\%, so no extra requirement on \mtl is made.
The lost-lepton component of the SM background in each SR, $N^\text{pred}_{\rm LL}$,
is estimated once again from the corresponding data  via a transfer factor, $T_\mathrm{LL}$, determined from simulation:
\begin{equation}
\label{equ:lostlep}
N^\text{pred}_{\rm LL} = N^\text{data}_{1\ell} T_\mathrm{LL},\quad T_\mathrm{LL} = \frac{N^\text{sim}_{0\ell}}{N^\text{sim}_{1\ell}},
\end{equation}
where $ N^\text{data}_{1\ell}$ is the observed event yield in the single-lepton CR.  The transfer factor $T_\mathrm{LL}$ accounts for effects related to lepton acceptance and efficiency.

The largest uncertainty in the lost-lepton estimate is, as in the previous analyses, due to the statistical uncertainty in the  event yields, ranging from 3 to 50\%, depending on the SR.
Contributions to the CRs from \zll~and diboson processes are subtracted using estimates from simulation, and a 50\% uncertainty is applied to this subtraction, which leads to an uncertainty of 3--10\% in the lost-lepton prediction.  The limited event counts in the simulation sample result in a 2--12\% uncertainty, while uncertainties related to discrepancies between the lepton selection efficiency in data and simulation give rise to a 3--4\% uncertainty in the final estimate.  An additional uncertainty of 7\% in the $\tau_{\rm h}$ component accounts for differences in isolation efficiency between muons and single-prong $\tau_{\rm h}$ decays, as determined from studies with simulated samples of \wjets and \ttbar events.
A systematic uncertainty of 5--10\% is found for the uncertainties in the \bq~tagging scale factors that are applied to the simulation for the differences in b tagging performance between data and simulation and the different b tagging working points.
Finally, we estimate a systematic uncertainty in the transfer factor to account for differences in the \ttbar and \wjets admixture in the search and control regions.   This results in a 20--30\% uncertainty in the final prediction.

\subsubsection{Estimation of the QCD background}
\label{sec:qcd}

The $\dphijonetwothree > 0.4$ requirement reduces the QCD contribution to a small fraction of the total background in all SRs for both compressed and noncompressed models.  We estimate this contribution for each SR by applying a transfer factor to the number of events observed in a CR enriched in QCD events. The CRs are obtained by inverting the $\dphijonetwothree$ requirement. The transfer factor, \TQCD, is measured in a sideband region in data with \met$\in[200,250]$\GeV and the same requirements on the other variables as in the SRs.  This factor is the ratio between the number of QCD events in the $\dphijonetwothree > 0.4$ and $\dphijonetwothree < 0.4$ subsets of this sideband region.  The estimated contribution of other SM processes (\ttbar,  \wjets, single top quark, and diboson production) based on simulated samples is subtracted from the event yields in the CR and each subset of the sideband.

 The transfer factor for the noncompressed regions does not vary significantly as a function of \HTonetwo\ and \mct. Therefore, we extract the value of \TQCD used for the noncompressed SRs from a sideband selected with an inclusive requirement on \HTonetwo\ and \mct to reduce the statistical uncertainty in the transfer factor. The transfer factors for the compressed SRs are obtained from sidebands that are subdivided according to the number of \bq-tagged jets into $\nb = 0$ and $\nb \geq 1$ regions, with the latter used to extract the QCD predictions for the $\nb = 1$ and $\nb \geq 2$ SRs.

The statistical uncertainties due to the limited number of events in the data CRs and the non-QCD simulated samples are propagated to the final QCD estimate, ranging from 10 to 100\%.  The main uncertainty in \TQCD also originates from the statistical uncertainty of the observed and simulated event yields in the sideband region. We assign additional uncertainties for differences in \bq~tagging efficiency between data and simulation and for the subtraction of the non-QCD background contribution in the sideband.  The total systematic uncertainty in the QCD prediction varies between 27\% and 76\% in the compressed SRs, but can be as large as 550\% in the noncompressed SRs due to the small event samples in the corresponding sideband in data.

\subsection{Results}
\label{sec:result}

\begin{figure}[!ht]
  \centering
    \includegraphics[width=0.9\columnwidth]{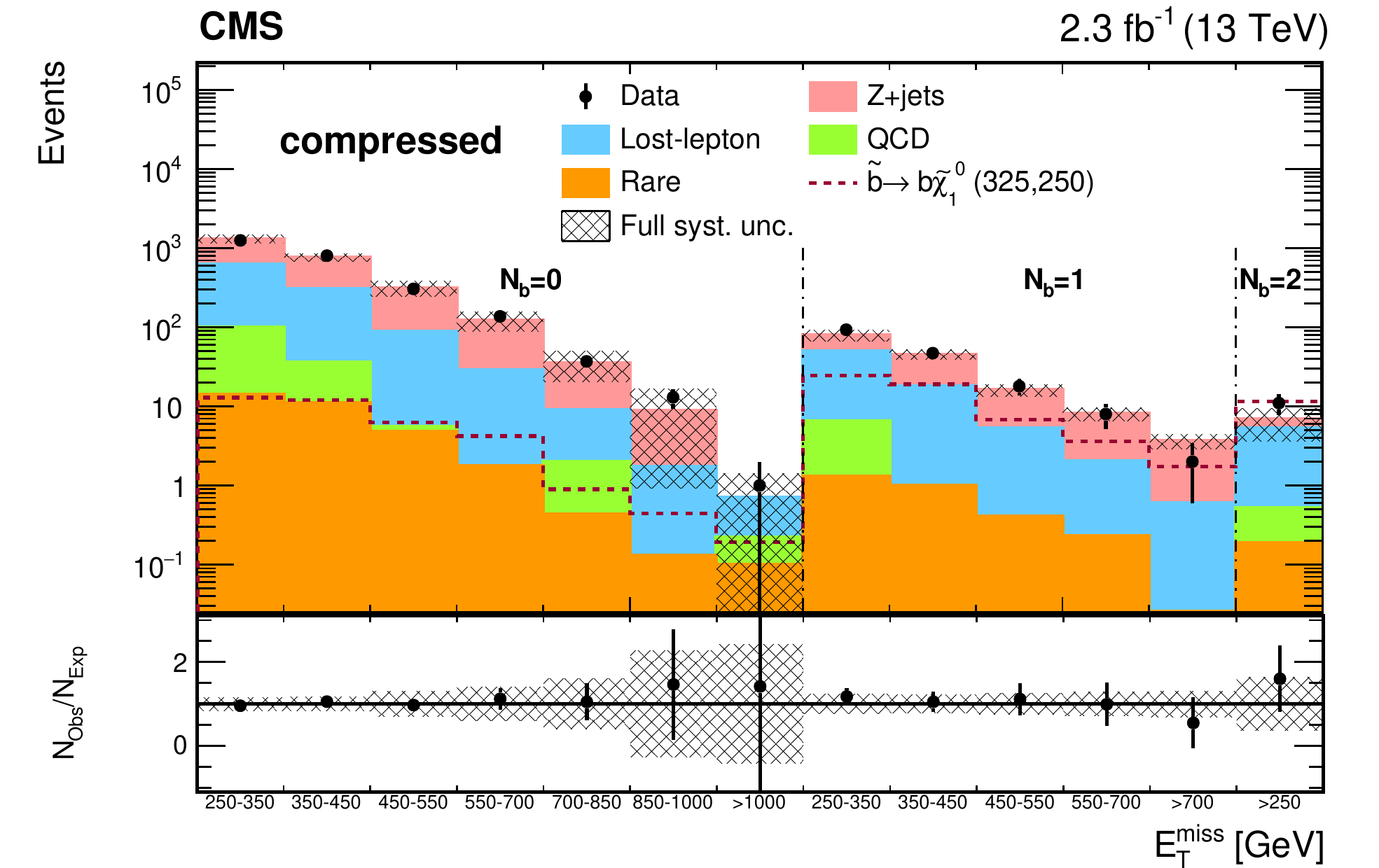}
    \includegraphics[width=0.9\columnwidth]{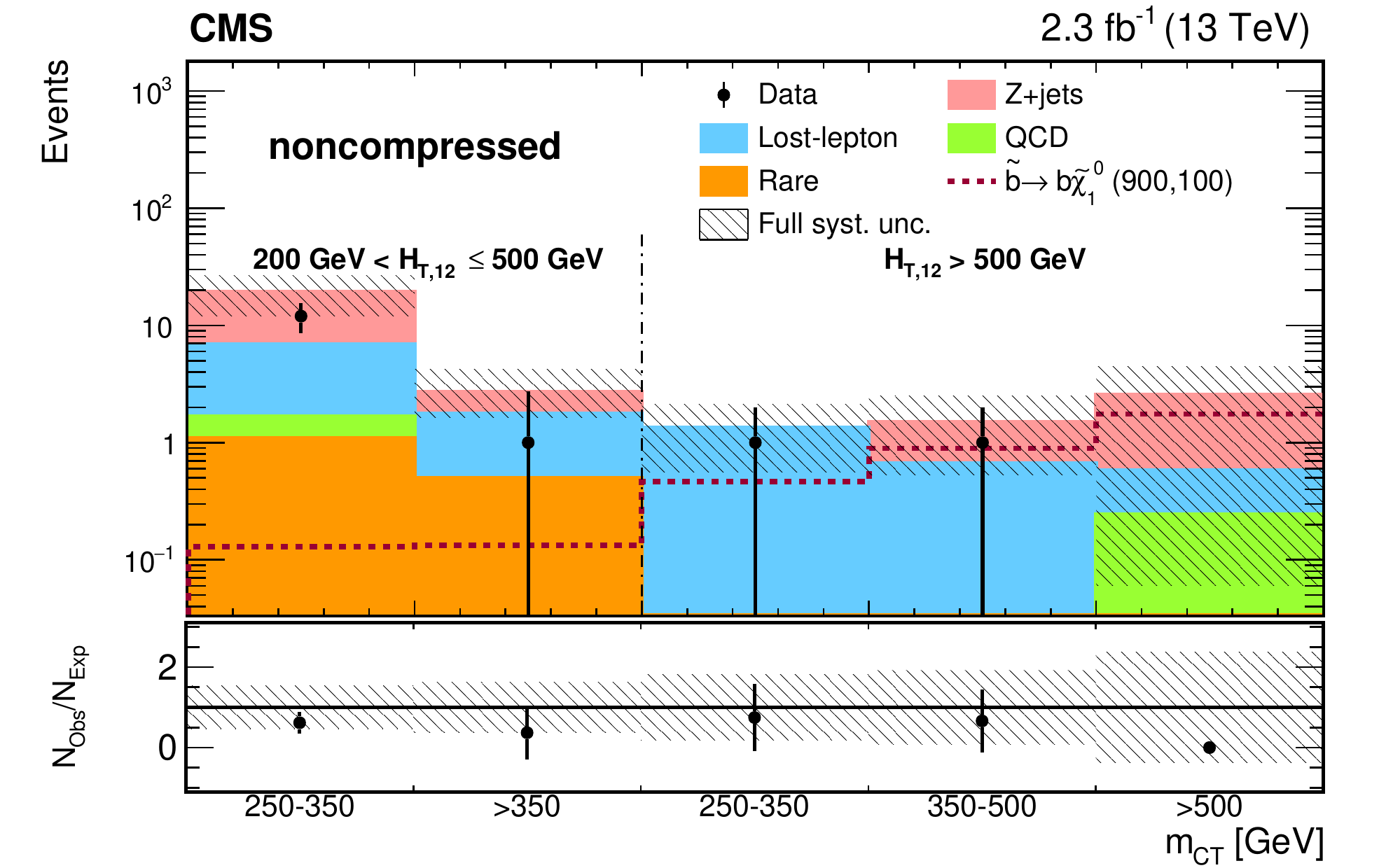}
    \caption{Observed events and estimated SM background and signal yields for the compressed (top) and noncompressed (bottom) SRs for the bottom squark search in the all-jet final state. The observed data yield is shown as black points and the total background predictions are shown in solid area. The SM background predictions shown do not include the effects of the maximum likelihood fit to the data. The bottom panel shows the ratio of data to the total background prediction in each search bin. Only statistical uncertainties are propagated to the ratio. }
    \label{fig:allPred}

\end{figure}

The expected SM background yields and the number of events observed in data are
summarized in Table~\ref{tab:pred} and shown in Fig.~\ref{fig:allPred}.  The observed yields agree well with the predicted SM background.

\begin{table*}[!ht]
\centering
\topcaption{Observed number of events and background prediction in the  different SRs for the $\sbottomq$ and $\stopq \to \cq \lsp$ searches.  The total uncertainty in the background predictions is also shown.
}
\begin{tabular}{ c  cccc c c}
\hline
 &  \znunu  &  Lost-lepton  &  QCD  &  Rare SM  &  Total SM  &  Data  \\
\hline
\mct [\GeVns{}] & \multicolumn{6}{c}{$200$\GeV$< \HTonetwo \leq 500$\GeV} \\
\hline
250--350 & 12.5$\pm$6.3 & 5.3$\pm$2.0 & 0.6$^{+3.3}_{-0.6}$ & 1.09$\pm$0.54 & 19.4$^{+7.4}_{-6.6}$ & 12 \\
$>$350 & 0.9$^{+1.1}_{-0.9}$ & 1.28$\pm$0.46 & $<$0.34 &  0.18$\pm$0.09  &  2.4$^{+1.3}_{-1.1}$ & 3 \\
\hline
\mct [\GeVns{}] &\multicolumn{6}{c}{$\HTonetwo >500$\GeV } \\
\hline
250--350 &$<$1.5  &1.34$\pm$0.78 &  $<$0.34   &$<$0.12 & 1.34$\pm$0.78 & 1 \\
350--500 & 0.84$\pm$0.94 & 0.67$\pm$0.35 &  $<$0.34  &$<$0.12 & 1.51$\pm$0.98 &1 \\
$>$500 & 2.0$\pm$1.6 & 0.34$\pm$0.20 & 0.2$^{+1.6}_{-0.2}$ & $<$0.12 &  2.3$^{+2.2}_{-1.6}$ & 0\\
\hline
\met [\GeVns{}] &\multicolumn{6}{c}{$\nb = 0$}  \\
\hline
250--350 & 680$\pm$78 & 530$\pm$120 & 86$\pm$25 & 14.2 $\pm$7.1 & 1310$\pm$150 & 1250 \\
350--450 & 454$\pm$63 & 270$\pm$64 & 24.9$\pm$8.8 & 11.0$\pm$5.5 & 760$\pm$89 & 802 \\
450--550 & 226$\pm$42 & 82$\pm$52 & 0.8$^{+2.7}_{-0.8}$ & 4.8$\pm$2.4 & 314 $\pm$67 & 305 \\
550--700 & 94$\pm$27 & 27 $\pm$21 &$<$0.95 & 1.75$\pm$0.87 & 122 $\pm$34 & 137 \\
700--850 & 26$\pm$14 & 7.0$\pm$6.1&  1.6$\pm$1.4 & 0.43$\pm$0.21&  35$\pm$15 & 37 \\
850--1000& 7.2$^{+7.6}_{-7.2}$ &1.6$^{+1.8}_{-1.6}$ &$<$0.95 &  0.13$\pm$0.06 & 7.3$^{+7.9}_{-7.3}$ & 13 \\
$>$1000 & $<$2.0 & 0.48$^{+0.51}_{-0.48}$ & 0.12$^{+0.53}_{-0.12}$ &0.11$\pm$0.05 & 0.71$^{+0.71}_{-0.52}$ & 1\\
\hline
\met [\GeVns{}] &  \multicolumn{6}{c}{$\nb = 1$}  \\
\hline
250--350 & 29.2$\pm$5.0 &  43$\pm$11 & 5.1$\pm$4.2 & 1.32$\pm$0.65 & 79$\pm$13 & 93 \\
350--450 & 27.7$\pm$4.7 & 17.1$\pm$4.9 & $<$0.47& 0.99$\pm$0.49& 45.8$\pm$6.8 & 47 \\
450--550 & 10.8$\pm$2.0 & 4.9$\pm$2.0 & $<$0.47& 0.41$\pm$0.20 & 16.2$\pm$2.8 & 18 \\
550--700 & 6.0$\pm$1.3 & 1.82$\pm$0.96 & $<$0.47& 0.23$\pm$0.11& 8.1$\pm$1.6 &  8\\
$ > 700 $  & 3.07$\pm$0.64 & 0.59$\pm$0.47 & $<$0.47& $<$0.12 &3.66$\pm$0.80 & 2 \\
\hline
\met [\GeVns{}] &  \multicolumn{6}{c}{$\nb = 2$}  \\
\hline
 $>$250  & 1.6$\pm$1.6 &  4.7$\pm$2.5 &  0.32$^{+0.40}_{-0.32}$ & 0.19$\pm$0.09 & 6.5$\pm$2.9 & 11 \\
\hline
\end{tabular}
\label{tab:pred}
\end{table*}

\section{Systematic uncertainties}
\label{sec:systematics}

Several categories of systematic uncertainties apply to all three analyses. These include uncertainties arising from the limited event counts in control samples, uncertainties related to the use of simulation in SM background predictions, and a 2.7\% uncertainty in integrated luminosity~\cite{CMS-PAS-LUM-15-001} that applies to the estimated signal yields and contributions from rare background processes that are taken directly from simulation, without the use of data control samples.

The limited number of simulated events surviving the stringent requirements on jets and \met in all three searches can lead to a significant statistical uncertainty in background predictions. In the case of background predictions that rely on simulation for accurate modelling of the relevant event kinematics, we assess theoretical uncertainties, primarily those associated with missing higher-order corrections, in the simulated samples by varying the renormalization and factorization scales up and down by a factor of two~\cite{Catani2003zt,Cacciari2003fi} and by variations of PDFs. The PDF uncertainties are defined by the standard deviation obtained from 100 variations of the NNPDF3.0~\cite{Ball:2014uwa} PDFs.  The uncertainties are then propagated to the final background estimates.

When the simulation of the detector response does not adequately describe the data, correction factors are applied to account for the observed discrepancies. Differences in the efficiencies for selecting isolated leptons between simulation and data are measured in \zll~events in the case of electrons and muons and in a \ttbar-enriched sample for hadronically decaying $\tau$ leptons. The observed deviations are accounted for in the form of corrections to the simulation, and the corresponding uncertainties are propagated to the predicted SM yields in the SRs. Correction factors and uncertainties based on measurements of \bq~tagging performance in data and simulation~\cite{CMS-PAS-BTV-15-001} are also applied. They are parameterized by jet kinematics and flavour. We also assess an uncertainty related to the modelling of additional interactions in the simulation. For the rare SM backgrounds with top quarks, predominantly from \ttbar production in association with a \Z boson, where the \Z boson decays to a pair of neutrinos, an extra uncertainty is estimated to account for the possible mismodelling of the top quark \pt spectrum.
The efficiency and misidentification rates for the top quark tagging algorithm are compared between data and simulation in CRs as a function of the key kinematic variables.  The correction factors are found not to be strongly dependent on the different kinematic variables.  The efficiency estimated in simulation agrees with the measured efficiency while the misidentification rate has to be corrected by 30\%.  Both correction factors have a 10\% uncertainty, estimated from the variations of the efficiency measurement.

All these uncertainties are propagated to the different signal and background estimates to which they apply.  The background predictions from control samples in data are affected through the transfer factors that are calculated from simulation corrected to reproduce data.  In general these uncertainties are subdominant and the uncertainty in the final background estimate is dominated by the statistical uncertainty of the data control sample.

For the signal samples differences between the fast simulation and the full \GEANTfour-based model are also taken into account.  Lepton selection efficiencies and \bq~tagging performance are found to be different in the fast simulation.  We derive appropriate corrections for the fast simulation and propagate the corresponding uncertainties to the predicted signal yields.  We also assess an additional uncertainty for the difference in \met resolution between the fast simulation and the full \GEANTfour-based model. This difference in \met resolution has the largest impact on signal models with small intrinsic \met, as is the case for compressed mass spectra.  The modelling of the ISR plays an important role in cases where the top squark and $\lsp$ masses are very similar.  The uncertainty is determined by comparing the simulated and observed \pt spectra of the system recoiling against the ISR jets in \ttbar events, using the method described in Ref.~\cite{CMS-STOP-lepton}. The effect is generally found to be small, although in scenarios with a compressed mass spectrum the effect can be as large as 30\%.

The uncertainties in the signal modelling are determined in each analysis for every SR.    The dominant uncertainties in the predicted signal yield arise from the size of the simulated samples in some of the SRs (1--100\%), jet energy scale corrections (1--50\%), $\PQb$ tagging efficiency corrections used to scale simulation to data (1--35\%), and ISR (1--30\%).  The largest uncertainties are in SRs that have small signal acceptance to a specific model.

The statistical uncertainties of the signal samples are uncorrelated, whereas all other signal systematic uncertainties are considered to be fully correlated among the different SRs and analyses.  Since the three analyses predict the backgrounds with different CRs, the treatment of systematic uncertainties is mostly uncorrelated among analyses, except for the estimates based on simulation. Here only the statistical component of the uncertainty is treated as uncorrelated. Systematic uncertainties due to jet energy scale corrections,  $\PQb$ tagging efficiency and selection efficiencies are treated as correlated among the different background estimates.

\section{Interpretation}
\label{sec:interpretation}
The data in all three searches are consistent with the background expected from SM processes. The results are interpreted as limits on supersymmetric particle masses in the context of simplified models~\cite{Simp,Simp1,Simp2,Simp3}
of top or bottom squark pair production.

Different decay modes are considered for top squark pair production.  For mass splittings $\Delta m$ larger than the \W~boson mass, we consider two decay modes for the top squark: to a top quark and a neutralino, or to a bottom quark and a chargino, where the chargino decays to an LSP. Scenarios with $\stopq\to \topq^{(*)}\lsp$ branching fractions of 50 or 100\% are considered.
The results of the top squark searches in the all-jet and single-lepton final states are combined for these interpretations. For $\Delta m$ smaller than the \W~boson mass, only the decay of top squarks to a charm quark and an LSP is considered in this paper.  For the pair production of bottom squarks, all bottom squarks are assumed to decay to a bottom quark and an LSP.

The signal yield is corrected for signal contamination of data CRs for each mass hypothesis and each analysis. Typical values are around 5--10\%, except for compressed mass spectra, where it can vary between 10 and 50\%.  The signal contamination is most significant for the top squark production models with a 100\% $\stopq\to \topq^{(*)}\lsp$ branching fraction, a light LSP, and $\Delta m$ close to the top quark mass.   The 95\% confidence level (CL) upper limits on SUSY production cross sections are calculated using a modified frequentist approach  with the CL$_\mathrm{S}$ criterion~\cite{Junk:1999kv,Read:2002hq} and asymptotic results for the test statistic~\cite{LHC-HCG,Cowan:2010js}.

The SRs and CRs for top squark searches in the all-jet and single-lepton final states  are mutually exclusive.  We combine the  results of the two searches, treating the systematic uncertainties assigned to the predicted signal and background yields as correlated or uncorrelated depending on the source, as detailed in Section~\ref{sec:systematics}.

\begin{figure}[htb]
\centering
\includegraphics[width=\cmsFigWidth]{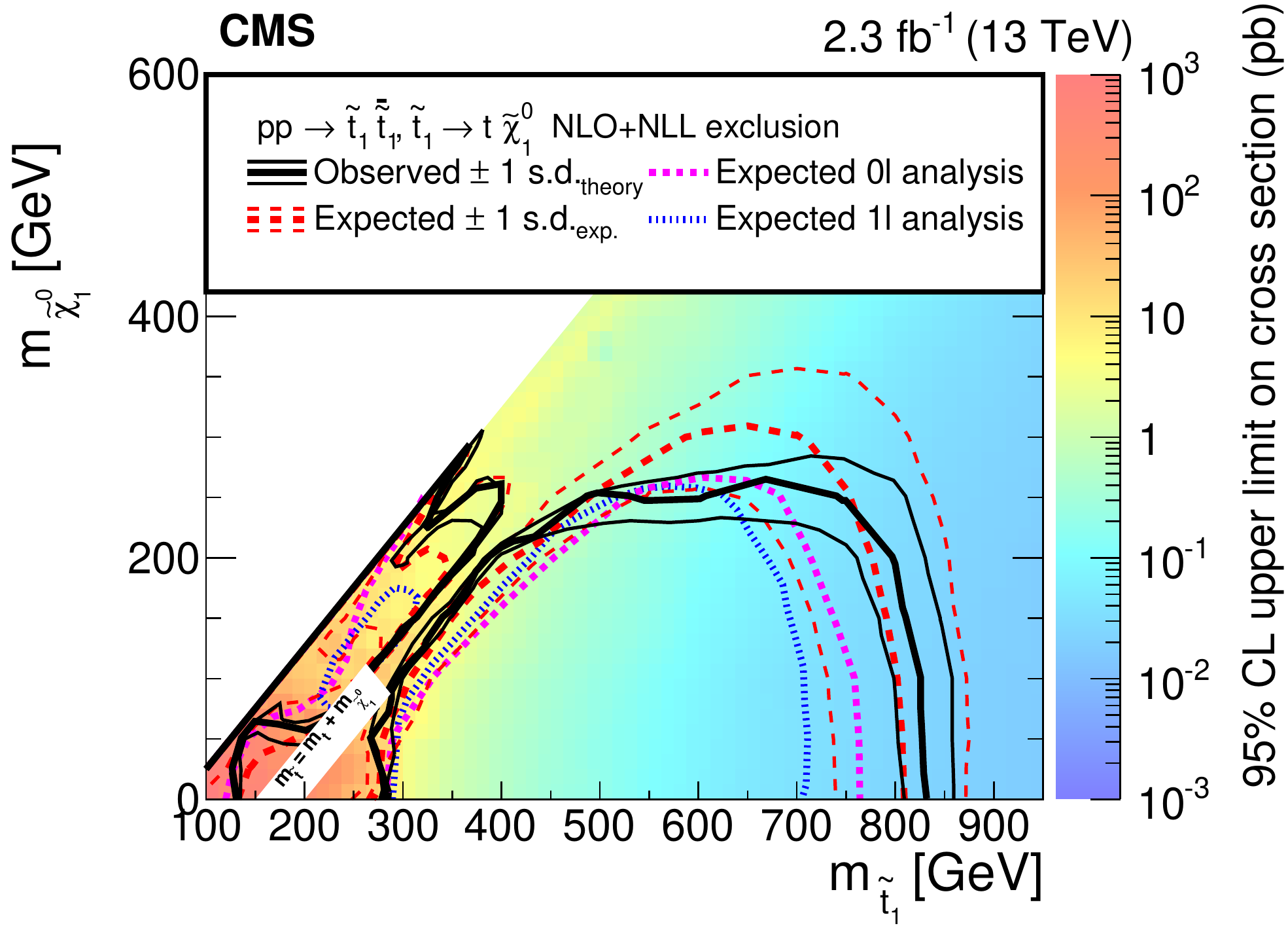}
\caption{\label{fig:limits:T2tt}Exclusion limits at 95\% CL for direct top squark pair production for the decay mode $\stopq\to\topq^{(*)}\lsp$.
 The interpretation is performed in the two-dimensional space of $m_{\stopq}$ vs. $m_{\lsp}$. The color indicates the 95\% CL upper
  limit on the product of cross section and branching fraction at each point in the $m_{\stopq}$-$m_{\lsp}$ plane.
  The regions enclosed by the thick black curves represent the observed exclusion at 95\% CL, while the dashed red lines indicate the expected limits at 95\% CL and their $\pm$1 s.d. experimental uncertainties.  The thin black lines show the impact of the $\pm$1 s.d. theoretical uncertainties in the signal cross section.  The magenta short-dashed curve and the blue dotted curve  show the expected limits for the analysis in the all-jet (Section~\ref{sec:stop0l}) and single-lepton (Section~\ref{sec:1lstop}) final states, respectively. The limits in the region near $\Delta m\approx m_{\topq}$ and low $\lsp$ mass are not shown due to the difficulty in modelling rapidly varying kinematics in this region.}
\end{figure}

Figure~\ref{fig:limits:T2tt} shows 95\% CL exclusion limits for $\Pp\Pp\to\stopq\stopqbar\to \topq^{(*)}\lsp\topqbar^{(*)}\lsp$, assuming the top quarks in the decay to be unpolarized, together with the upper limit at 95\% CL on the excluded signal cross section. All top squarks are assumed to decay to a top quark and an LSP.  For $\Delta m<m_{\topq}$ the signal samples assume a three-body decay without an off-shell top quark as intermediate particle. The expected exclusion is given by the dashed red line, with the  one standard deviation (s.d.) experimental uncertainty.  The observed exclusion curve is shown as a solid black line together with the 1 s.d. uncertainty in the theoretical cross section.  We do not interpret in the region near $\Delta m \approx m_{\topq}$ when \lsp is very light because of the difficulty in modelling rapidly varying kinematics in this region. In this region an indirect search for top squark pair production can be performed by looking for a small excess in the measured \ttbar cross section compared to the SM expectation~\cite{TOP-13-004,atlas-stop1l-2015}.  We exclude top squark masses from 280\GeV to 830\GeV for a massless LSP and LSP masses up to 260\GeV for 675\GeV top squarks.  At 8 TeV top squark masses were excluded up to 780\GeV for a massless LSP~\cite{stop0l_8TeV}. For models with heavy top squarks and light LSPs, the sensitivity is driven by the top squark analysis in the all-jet final state of Section~\ref{sec:stop0l}, which is more sensitive than the single-lepton analysis (Section~\ref{sec:1lstop}) because of the larger acceptance for signal.  The combination extends the expected reach in top squark mass by about 45\GeV.  When the LSP is heavier, the cleaner search in the single-lepton final state becomes more important.  Both analyses have similar sensitivity in this area of parameter space, and combining them extends the reach in LSP mass by about 30\GeV.

\begin{figure}[htb]
\centering
\includegraphics[width=\cmsFigWidth]{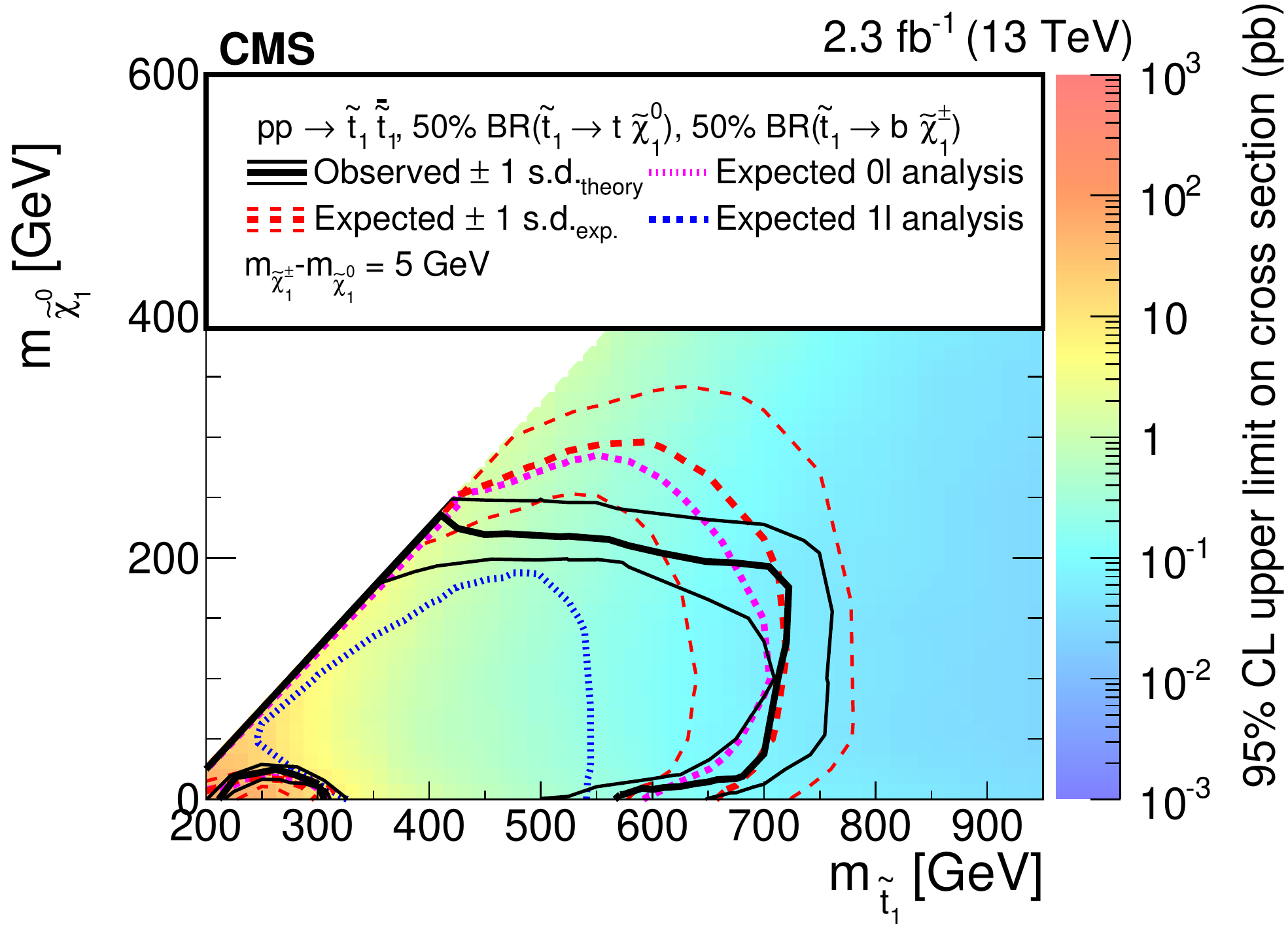}
\caption{\label{fig:limits:T2tb}Exclusion limits at 95\% CL for direct top squark pair production assuming equal branching fractions for the decays $\stopq\to\topq^{(*)}\lsp$ and $\stopq\to \bq\chipmone$.
 The interpretation is performed in the two-dimensional space of $m_{\stopq}$ vs. $m_{\lsp}$.  The chargino is considered to be nearly mass-degenerate with the LSP ($m_{\chipmone} = m_{\lsp} + 5$\GeV). The caption of Fig.~\ref{fig:limits:T2tt} explains the use of lines and colors in detail.}
\end{figure}

Figure~\ref{fig:limits:T2tb} shows the 95\% CL exclusion limits for $\stopq\stopqbar$ production, assuming equal probabilities for the decay modes $\stopq\to \topq^{(*)}\lsp$ and $\stopq\to \bq\chipmone$.  The chargino in the latter mode decays to a \W~boson and an LSP.  In this model, the chargino is considered to be nearly mass-degenerate with the LSP ($m_{\chipmone} = m_{\lsp} + 5$\GeV).  The \W~boson decay products originating from the chargino decay are very soft because of the small mass splitting, and might not be detectable.  For intermediate LSP masses, top squark masses are probed up to 725\GeV.
 The LSP masses up to 210\GeV are probed for a top squark mass of around 500\GeV.  Here, the single-lepton analysis does not contribute much to the combination because of the larger acceptance in the all-jet final state, except at low LSP masses.  In most of the mass parameter space the combination reaches $\approx$ 15\GeV higher than the analysis in the all-jet final state.

\begin{figure}[htb]
\centering
\includegraphics[width=\cmsFigWidth]{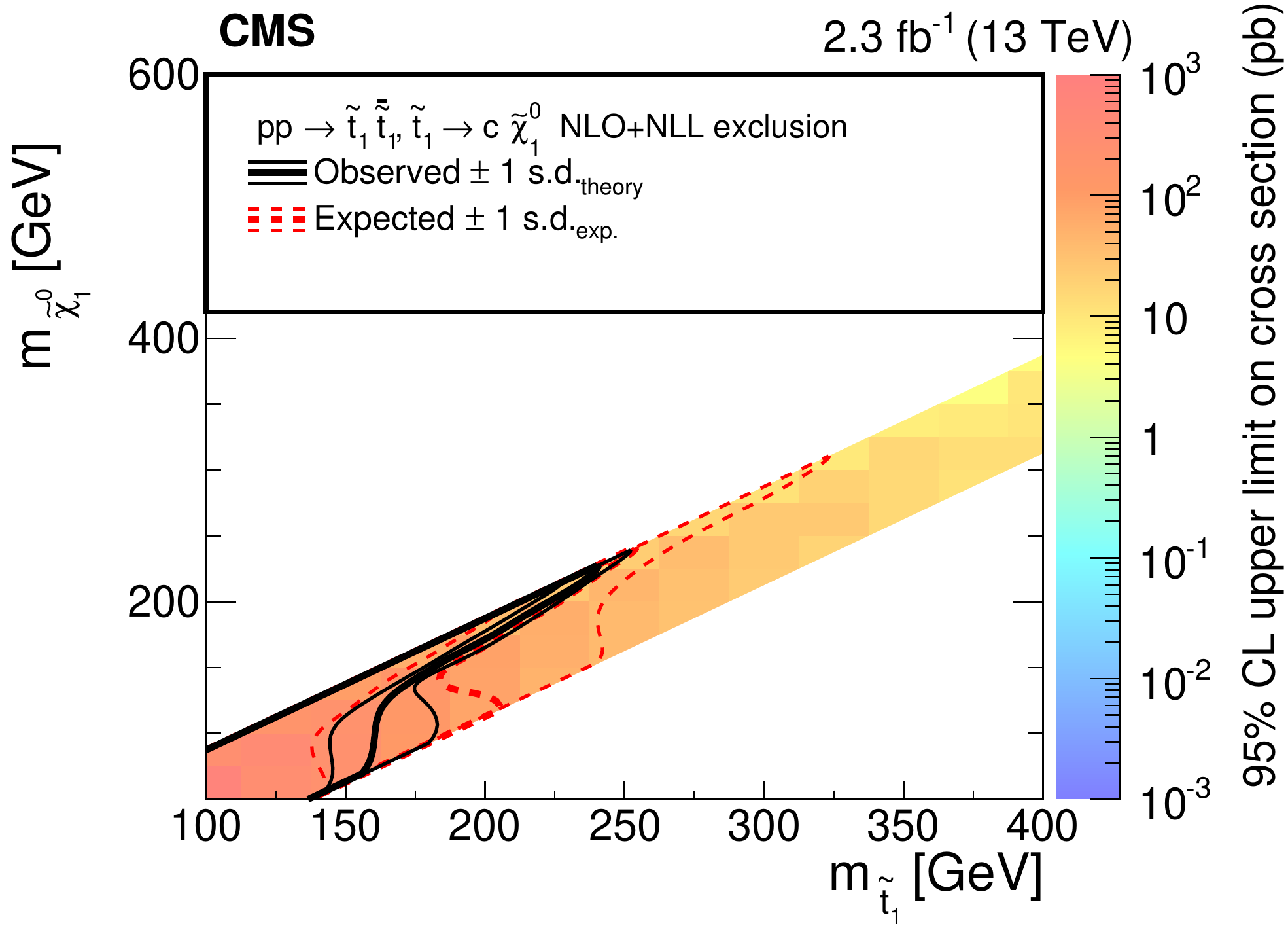}
\caption{\label{fig:limits:T2cc}Exclusion limits at 95\% CL for direct top squark pair production with decay $\stopq\to \cq\lsp$ using the compressed SRs of the bottom squark analysis (Section~\ref{sec:sbottom}).
 The interpretation is done in the two-dimensional space of $m_{\stopq}$ vs. $m_{\lsp}$. The caption of Fig.~\ref{fig:limits:T2tt} explains the use of lines and colors in detail.}
\end{figure}

The compressed SRs from the bottom squark analysis in the all-jet final state (Section~\ref{sec:sbottom}) are used to set upper limits on the top squark cross sections when the mass splitting between the top squark and the LSP is smaller than the mass of the \W~boson. Figure~\ref{fig:limits:T2cc} shows the expected and observed 95\% CL upper limits on the top squark cross sections in the $m_{\stopq}$-$m_{\lsp}$ plane assuming the top squark always decays to a charm quark and an LSP.   Top squarks with masses below 240\GeV are probed in this model, when the mass splitting between the top squark and the LSP is close to 10\GeV.  At 8\TeV top squark masses up to 270\GeV were probed for the same $\Delta$m~\cite{atlas-stop1l-2015}.

\begin{figure}[htb]
\centering
\includegraphics[width=\cmsFigWidth]{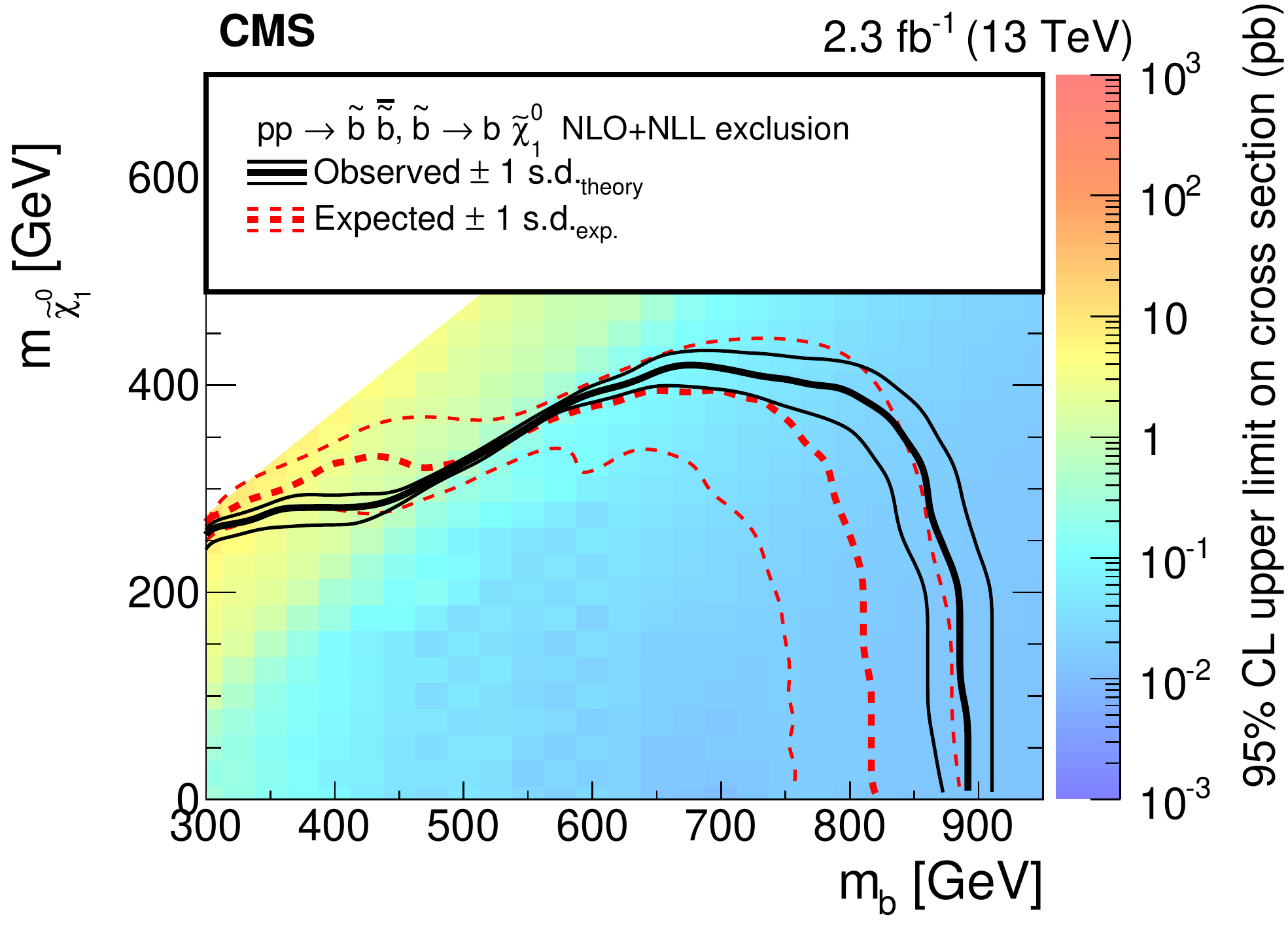}
\caption{\label{fig:limits:T2bb}Exclusion limits at 95\% CL for direct bottom squark pair production with decay $\sbottomq\to \bq\lsp$.
 The interpretation is performed in the two-dimensional space of $m_{\sbottomq}$ vs. $m_{\lsp}$ using the results of the bottom squark analysis (Section~\ref{sec:sbottom}). The caption of Fig.~\ref{fig:limits:T2tt} explains the use of lines
and colors in detail.}
\end{figure}

Figure \ref{fig:limits:T2bb} shows the expected and observed 95\% CL upper limits on the bottom squark cross sections in the $m_{\sbottomq}$-$m_{\lsp}$ plane using both the compressed and noncompressed SRs of the bottom squark analysis. We probe bottom squark masses up to 890\GeV for small LSP masses.  With 8 TeV data bottom squark masses below 650\GeV were excluded.~\cite{atlas-stop1l-2015,stop8TeV}.

\section{Summary}
\label{sec:summary}
Results are presented from three complementary searches for top or bottom squark-antisquark pairs in data collected with the CMS detector in proton-proton collisions at a centre-of-mass energy of 13\TeV, corresponding to an integrated luminosity of 2.3\fbinv . The search for top squarks is carried out in the all-jet and single-lepton final states, which are combined for the final result. A second search in all-jet events is designed for bottom squark pairs and for top squarks decaying to charm quarks through a flavour changing neutral current process. No statistically significant excess of events is observed above the expected standard model background, and exclusion limits are set at 95\% confidence level in the context of simplified models of direct top and bottom squark pair production.   Limits for top squark masses of 830\GeV are established for a massless lightest supersymmetric particle (LSP), and for LSP masses up to 260\GeV for a 675\GeV top squark mass, when  all top squarks are assumed to decay to a top quark and an LSP.  When the top squarks can also decay to a bottom quark and a chargino, this reach is reduced.  Assuming a mass splitting between the top squark and the LSP close to 10\GeV, and top squarks that decay to a charm quark and an LSP, top squark mass limits up to 240\GeV are established.  Finally, bottom squark mass limits up to 890\GeV are established for small LSP masses.  The results extend the reach with respect to previous limits obtained from LHC Run 1 data in most of the parameter space.
\begin{acknowledgments}
\hyphenation{Bundes-ministerium Forschungs-gemeinschaft Forschungs-zentren Rachada-pisek} We congratulate our colleagues in the CERN accelerator departments for the excellent performance of the LHC and thank the technical and administrative staffs at CERN and at other CMS institutes for their contributions to the success of the CMS effort. In addition, we gratefully acknowledge the computing centres and personnel of the Worldwide LHC Computing Grid for delivering so effectively the computing infrastructure essential to our analyses. Finally, we acknowledge the enduring support for the construction and operation of the LHC and the CMS detector provided by the following funding agencies: the Austrian Federal Ministry of Science, Research and Economy and the Austrian Science Fund; the Belgian Fonds de la Recherche Scientifique, and Fonds voor Wetenschappelijk Onderzoek; the Brazilian Funding Agencies (CNPq, CAPES, FAPERJ, and FAPESP); the Bulgarian Ministry of Education and Science; CERN; the Chinese Academy of Sciences, Ministry of Science and Technology, and National Natural Science Foundation of China; the Colombian Funding Agency (COLCIENCIAS); the Croatian Ministry of Science, Education and Sport, and the Croatian Science Foundation; the Research Promotion Foundation, Cyprus; the Secretariat for Higher Education, Science, Technology and Innovation, Ecuador; the Ministry of Education and Research, Estonian Research Council via IUT23-4 and IUT23-6 and European Regional Development Fund, Estonia; the Academy of Finland, Finnish Ministry of Education and Culture, and Helsinki Institute of Physics; the Institut National de Physique Nucl\'eaire et de Physique des Particules~/~CNRS, and Commissariat \`a l'\'Energie Atomique et aux \'Energies Alternatives~/~CEA, France; the Bundesministerium f\"ur Bildung und Forschung, Deutsche Forschungsgemeinschaft, and Helmholtz-Gemeinschaft Deutscher Forschungszentren, Germany; the General Secretariat for Research and Technology, Greece; the National Scientific Research Foundation, and National Innovation Office, Hungary; the Department of Atomic Energy and the Department of Science and Technology, India; the Institute for Studies in Theoretical Physics and Mathematics, Iran; the Science Foundation, Ireland; the Istituto Nazionale di Fisica Nucleare, Italy; the Ministry of Science, ICT and Future Planning, and National Research Foundation (NRF), Republic of Korea; the Lithuanian Academy of Sciences; the Ministry of Education, and University of Malaya (Malaysia); the Mexican Funding Agencies (BUAP, CINVESTAV, CONACYT, LNS, SEP, and UASLP-FAI); the Ministry of Business, Innovation and Employment, New Zealand; the Pakistan Atomic Energy Commission; the Ministry of Science and Higher Education and the National Science Centre, Poland; the Funda\c{c}\~ao para a Ci\^encia e a Tecnologia, Portugal; JINR, Dubna; the Ministry of Education and Science of the Russian Federation, the Federal Agency of Atomic Energy of the Russian Federation, Russian Academy of Sciences, the Russian Foundation for Basic Research and the Russian Competitiveness Program of NRNU MEPhI (M.H.U.); the Ministry of Education, Science and Technological Development of Serbia; the Secretar\'{\i}a de Estado de Investigaci\'on, Desarrollo e Innovaci\'on and Programa Consolider-Ingenio 2010, Spain; the Swiss Funding Agencies (ETH Board, ETH Zurich, PSI, SNF, UniZH, Canton Zurich, and SER); the Ministry of Science and Technology, Taipei; the Thailand Center of Excellence in Physics, the Institute for the Promotion of Teaching Science and Technology of Thailand, Special Task Force for Activating Research and the National Science and Technology Development Agency of Thailand; the Scientific and Technical Research Council of Turkey, and Turkish Atomic Energy Authority; the National Academy of Sciences of Ukraine, and State Fund for Fundamental Researches, Ukraine; the Science and Technology Facilities Council, UK; the US Department of Energy, and the US National Science Foundation.

Individuals have received support from the Marie-Curie programme and the European Research Council and EPLANET (European Union); the Leventis Foundation; the A. P. Sloan Foundation; the Alexander von Humboldt Foundation; the Belgian Federal Science Policy Office; the Fonds pour la Formation \`a la Recherche dans l'Industrie et dans l'Agriculture (FRIA-Belgium); the Agentschap voor Innovatie door Wetenschap en Technologie (IWT-Belgium); the Ministry of Education, Youth and Sports (MEYS) of the Czech Republic; the Council of Science and Industrial Research, India; the HOMING PLUS programme of the Foundation for Polish Science, cofinanced from European Union, Regional Development Fund, the Mobility Plus programme of the Ministry of Science and Higher Education, the National Science Center (Poland), contracts Harmonia 2014/14/M/ST2/00428, Opus 2014/13/B/ST2/02543, 2014/15/B/ST2/03998, and 2015/19/B/ST2/02861, Sonata-bis 2012/07/E/ST2/01406; the Thalis and Aristeia programmes cofinanced by EU-ESF and the Greek NSRF; the National Priorities Research Program by Qatar National Research Fund; the Programa Clar\'in-COFUND del Principado de Asturias; the Rachadapisek Sompot Fund for Postdoctoral Fellowship, Chulalongkorn University and the Chulalongkorn Academic into Its 2nd Century Project Advancement Project (Thailand); and the Welch Foundation, contract C-1845.

\end{acknowledgments}

\bibliography{auto_generated}

\cleardoublepage \appendix\section{The CMS Collaboration \label{app:collab}}\begin{sloppypar}\hyphenpenalty=5000\widowpenalty=500\clubpenalty=5000\textbf{Yerevan Physics Institute,  Yerevan,  Armenia}\\*[0pt]
A.M.~Sirunyan, A.~Tumasyan
\vskip\cmsinstskip
\textbf{Institut f\"{u}r Hochenergiephysik,  Wien,  Austria}\\*[0pt]
W.~Adam, E.~Asilar, T.~Bergauer, J.~Brandstetter, E.~Brondolin, M.~Dragicevic, J.~Er\"{o}, M.~Flechl, M.~Friedl, R.~Fr\"{u}hwirth\cmsAuthorMark{1}, V.M.~Ghete, C.~Hartl, N.~H\"{o}rmann, J.~Hrubec, M.~Jeitler\cmsAuthorMark{1}, A.~K\"{o}nig, I.~Kr\"{a}tschmer, D.~Liko, T.~Matsushita, I.~Mikulec, D.~Rabady, N.~Rad, B.~Rahbaran, H.~Rohringer, J.~Schieck\cmsAuthorMark{1}, J.~Strauss, W.~Waltenberger, C.-E.~Wulz\cmsAuthorMark{1}
% ...
\vskip\cmsinstskip
% ...
1:~~Also at Vienna University of Technology, Vienna, Austria\\
% ...

\end{sloppypar}
\end{document}